\documentclass[10pt,journal,compsoc]{IEEEtran}

\usepackage{amsmath,amssymb,amsfonts}
\usepackage{algorithmic}
\usepackage{graphicx}
\usepackage{textcomp}
\usepackage{xspace}
\usepackage{xcolor}
\usepackage{fontawesome}
\usepackage{multirow}
\usepackage{rotating}
\usepackage{tikz}
\usepackage{soul}
\usepackage{makecell}
\usepackage{url}
\usepackage{enumitem}
\usepackage{booktabs}
\usepackage{fontawesome}
\usepackage{pifont}
\usepackage{amssymb} 
\usepackage[colorlinks=true, allcolors=black]{hyperref}
\usepackage{tcolorbox}
\usepackage{colortbl}
\usepackage[caption=false]{subfig}
\usepackage[export]{adjustbox}
\usepackage{ifthen}
\usepackage[normalem]{ulem}
\usepackage{etaremune}

\ifCLASSOPTIONcompsoc
  \usepackage[nocompress]{cite}
\else
  \use package{cite}
\fi

\newcommand{\ie}{\emph{i.e.,}\xspace}
\newcommand{\eg}{\emph{e.g.,}\xspace}
\newcommand{\etc}{etc.\xspace}
\newcommand{\etal}{\emph{et~al.}\xspace}
\newcommand{\secref}[1]{Section~\ref{#1}\xspace}

\newcommand{\figref}[1]{Fig.~\ref{#1}\xspace}

\newcommand{\tabref}[1]{Table~\ref{#1}\xspace}

\newboolean{showcomments}
\setboolean{showcomments}{true}

\ifthenelse{\boolean{showcomments}}
  {\newcommand{\nb}[2]{
    \fbox{\bfseries\sffamily\scriptsize#1}
    {\sf\small$\blacktriangleright$\textit{#2}$\blacktriangleleft$}
   }
  }
  {\newcommand{\nb}[2]{}
  }

\newcommand{\developers}{nine\xspace}
\newcommand{\summaries}{$\sim$1.2k\xspace}
\newcommand{\totallyAssessedFunctions}{2,686\xspace}
\newcommand{\totallyAssessedFunctionsJava}{1,405\xspace}
\newcommand{\totallyAssessedPythonFunctions}{1,281\xspace}
\newcommand{\LLMs}{DeepSeek Coder 1.3B, 6.7B, and 33B \cite{guo:2024}, CodeLlama 7B, 13B, and 34B \cite{codellama}, GPT-3.5-turbo and GPT-4-turbo \cite{chatgpt}\xspace}
\newcommand{\CG}{\emph{code generation}\xspace}
\newcommand{\CS}{\emph{code summarization}\xspace}

\begin{document}

\title{On the Effectiveness of LLM-as-a-judge\\for Code Generation and Summarization}

\author{Giuseppe~Crupi,
	Rosalia~Tufano,
	Alejandro~Velasco,\\
	Antonio~Mastropaolo,
	Denys~Poshyvanyk,
	and~Gabriele~Bavota
	\IEEEcompsocitemizethanks{\IEEEcompsocthanksitem G. Crupi is with SEART @  Software Institute, Universit\`a della Svizzera italiana, Switzerland. \protect\\
		E-mail: giuseppe.crupi@usi.ch
		\IEEEcompsocthanksitem R. Tufano is with SEART @  Software Institute, Universit\`a della Svizzera italiana, Switzerland. \protect\\
		E-mail: rosalia.tufano@usi.ch
		\IEEEcompsocthanksitem A. Velasco is with W\&M. \protect\\
		E-mail: svelascodimate@wm.edu
		\IEEEcompsocthanksitem A. Mastropaolo is with  W\&M. \protect\\
		E-mail: amastropaolo@wm.edu
		\IEEEcompsocthanksitem D. Poshyvanyk is with W\&M. \protect\\
		E-mail: denys@cs.wm.edu
		\IEEEcompsocthanksitem G. Bavota is with SEART @  Software Institute, Universit\`a della Svizzera italiana, Switzerland. \protect\\
		E-mail: gabriele.bavota@usi.ch}
}

\markboth{Journal of \LaTeX\ Class Files,~Vol.~xx, No.~x, Month~xxxx}%
{Crupi \MakeLowercase{\etal}: On the Effectiveness of LLM-as-a-judge\\for Code Generation and Summarization}
\IEEEtitleabstractindextext{%
\begin{abstract}
Large Language Models (LLMs) have been recently exploited as judges for complex natural language processing tasks, such as Q\&A (Question \& Answer). The basic idea is to delegate to an LLM the assessment of the ``quality'' of the output provided by an automated technique (often another LLM) for tasks for which: (i) quantitative metrics would only tell part of the story, and; (ii) a large-scale human-based evaluation would be too expensive. LLMs-as-a-judge, if proven effective for a specific task, can also unlock new possibilities for automation, with several LLMs proposing a solution for a given instance of the task (\eg an answer to a question) and others judging and deciding what is the best output to show the user. We study the effectiveness of LLMs-as-a-judge for two code-related tasks, namely \CG and \CS. The rationale for choosing these tasks is two-fold. First, quantitative metrics are usually not enough for the assessment of code summarizers/generators. For example, it is well documented that metrics such as BLEU are quite weak proxies for the quality of the generated summaries. Second, even state-of-the-art techniques still struggle with handling complex instances of these tasks (\eg summarizing a quite long / complex function), making them good candidates for benefiting from more advanced solutions envisioning collaboration among LLMs. For \CG, we check whether eight LLMs are able to judge the correctness of \totallyAssessedFunctionsJava Java methods and \totallyAssessedPythonFunctions Python functions generated by the same LLMs or implemented by humans. For \CS, we compare the judgment of five LLMs to those provided by \developers humans for \summaries summaries, related to both Java and Python functions.  Our findings show that GPT-4-turbo is the best LLM in terms of judging capabilities for both tasks, with ``smaller'' LLMs featuring tens of billions parameters not being able to cope with judging tasks. However, even the best-performing LLM frequently misjudges the correctness of the code and summary quality.
\end{abstract}

}

\maketitle

\IEEEdisplaynontitleabstractindextext
\IEEEpeerreviewmaketitle

\section{Introduction} \label{sec:intro}

Large Language Models (LLMs) have been exploited in software engineering (SE) to (partially) automate a variety of tasks, including code generation \cite{Ciniselli:tse2022,Xueying:icse2024}, clone detection \cite{10.1145/2970276.2970326}, code summarization \cite{LeClair:icpc2020,Toufique:ase2022}, code review \cite{Tufano:icse2022,Zhiyu:fse2022}, bug fixing \cite{Matthew:fse2023,Guoyang:2023,8668043} and others \cite{Watson:tosem2022}. The wide adoption of LLMs in SE resulted in a shift in focus from classification (\eg defect detection, requirements classification) to generative problems requiring the synthesis of text (\eg code summarization) and/or code (\eg code generation). Generative problems, in addition to being harder to automate, also bring with them challenges related to the evaluation of automated solutions. For example, assessing if a natural language text is a good summary for a given code would require developers' judgment. Since human-based studies are expensive and difficult to scale up, researchers frequently employ quantitative metrics such as BLEU \cite{papineni2002bleu}, ROUGE \cite{lin2004rouge}, and METEOR \cite{meteor} as proxies for the ``quality'' of the generated summaries. These metrics are based on the idea that the more similar the generated summary is to a \emph{reference} summary (usually being the comment written by the developers for the code provided as input), the higher its quality. However, recent works in the literature \cite{roy:fse2021,haque:icpc2022,jin2024simllm,mastropaolo:icse2024} showed the shortcomings of these metrics when applied to code summarization: First, the reference summary, usually extracted from software repositories, may be of low quality, thus not being a good ``oracle'' to compare with. Second, the generated summary may be completely different from the reference one, while still being of high quality. In short, the automatic assessment of code summarization techniques and, more generally, of approaches automating software-related generative tasks, is still an open research avenue. 

A helping hand to address this challenge may come from LLMs themselves. Indeed, Lianmin \etal \cite{Lianmin:2023} proposed the idea of LLM-as-a-judge in the natural language processing (NLP) field: The LLM is prompted with instructions asking it to evaluate the output produced by other techniques (often other LLMs) according to specific criteria. Recent works also pioneered the usage of LLM-as-a-judge for assessing the correctness of an implementation given functional requirements described in natural language \cite{zhuo-2024-ice,tong-zhang-2024-codejudge}. While these works report some preliminary promising findings, they also show that LLMs struggle when dealing with the judgement of more complex implementations. In this paper, we build on top of these works, experimenting with more recent LLMs and with two code-related tasks. The first is the already presented \CS. The second is \CG, namely the automated implementation of a functionality starting from its natural language intent (\ie textual description)\footnote{Same task investigated in \cite{zhuo-2024-ice,tong-zhang-2024-codejudge}}. We focus on these two tasks since (i) they have been the subject of works proposing AI-based techniques for their automation \cite{Watson:tosem2022}, and (ii) due to their generative nature, they pose major challenges in the assessment of techniques automating them. For \CS, we already discussed these challenges. Concerning automatically generated code, researchers usually exploit two evaluation strategies. The first relies on code similarity metrics (\eg CrystalBLEU \cite{Aryaz:ase2022}) indicating how close the generated code is to a reference one. These metrics, however, are not a good proxy for the ``correctness'' of the implemented code. Indeed, as observed for summaries, the same function may be implemented in different ways, thus resulting in misjudging correct code being different from the expected target. The second strategy is aimed at exploiting benchmarks such as HumanEval \cite{humaneval} or CoderEval \cite{yu2024codereval}, presenting code generation problems featuring (i) a textual description of the code to implement, and (ii) a test suite to check the correctness of the generated code. Using these benchmarks, it is possible to compute metrics such as $pass@k$, which assesses the percentage of cases in which the LLM was able to generate a solution passing the tests using a maximum of $k$ attempts. For example, $pass@1$=50\% indicates that the LLM managed to generate a ``correct'' solution (according to the tests) for 50\% of the code generation problems using only a single attempt for each problem. While extremely valuable, these benchmarks are usually quite limited in dimension (\eg HumanEval features 160 Python code generation problems), resulting in a $pass@k$ value which only reflects the LLM's performance on a small set of coding problems, which does not really promote generalizability. Also, as we will demonstrate later, the accompanying test suites cannot always be trusted, even when assessing some trivial cases. For example, we found that for 9 of the 230 CoderEval Java problems \cite{yu2024codereval} a trivial empty function (\ie a function only having the signature) passes the associated tests, thus possibly inflating the $pass@k$ score in an empirical evaluation. Having LLMs able to reliably judge the correctness of automatically-generated code would allow to substantially scale-up the evaluations performed in \CG, even in scenarios in which tests are not available (or not completely reliable).

For \CG, we instructed eight LLMs (\ie \LLMs) to generate functions for the CoderEval benchmark \cite{yu2024codereval}, which features 230 Java and 230 Python code generation problems. Then, the LLMs judged the correctness of each other's solution as well as of functions written by humans, for a total of \totallyAssessedFunctions assessments per model. During the judgement task, we only provided the LLMs with (i) the textual description of the code to implement, and (ii) a function to be judged, either generated by the LLM or written by humans. We do not also provide an example of reference (correct) implementation for the code to implement. While this would substantially help the model in the judging task, it does not represent a real usage scenario in which \eg we want to use the LLM in the context of automated code review to assess the correctness of a given implementation (for which a reference implementation is not available). We verified the appropriateness of LLMs' assessment by running the generated functions on the test suites provided for each problem, checking if there is a relationship between a positive correctness assessment made by the LLMs and the test results (pass/fail). 

For \CS we compare the judgments provided by five of the eight LLMs (since three were clearly not able to perform the task) for 1,163 summaries ($\sim$50\% related to Java $\sim$50\% to Python functions) with judgments expressed by humans for those same summaries. Human judgments have been performed by a total of \developers humans (with \CS background), ensuring that each summary has been independently assessed by three humans. In the \CS study, the judged summaries include both summaries automatically generated by the same judged LLMs, as well as summaries written by humans.

We show that GPT-4-turbo is the best judge for both tasks, with smaller LLMs substantially struggling (\eg not being able to execute the task or presenting completely wrong judgments). GPT-4-turbo, while the best, still frequently misjudges the correctness of the code, especially lacking in the identification of wrong implementations, which are misjudged as correct in 50\% of the cases. GPT-4-turbo works better for assessing code summary quality, with moderate agreement with human judgments. 
\section{Study Design} \label{sec:design}
The \emph{goal} of our study is to assess the effectiveness of LLMs-as-a-judge for software-related tasks.
In particular, we formulate the following research question (RQ): \emph{To what extent can LLMs act as a judge for code generation and summarization?}

\subsection{Context Selection: LLMs}
\label{sub:llms}
We use eight LLMs (\ie \LLMs) as judges. Since the LLMs must be able to judge generated code and code summaries, we only use LLMs that have been trained on a corpus including source code. Also, we consider LLMs of different sizes (\ie number of trainable parameters) since the larger the language model, the higher its inference (judgment) cost. The latter may become a factor in deciding whether to use an LLM as a judge since the cost may be excessive to judge \eg millions of generated functions. In the following, we briefly describe each family of LLMs we consider. For all models, we consider their ``\emph{instruct}'' variants, that is, those also trained to understand instructions expressed in natural language.

\textbf{DeepSeek Coder \cite{guo:2024}.} DeepSeek Coder is a range of LLMs of different sizes, all trained from scratch on a corpus of 2 trillion tokens (87\% code, 10\% English and 3\% Chinese text). The \emph{instruct} versions of the model have then been fine-tuned on a dataset featuring human instructions with the following dialogues. We use DeepSeek Coder versions that have 1.3B, 6.7B, and 33B parameters. We generate judgments by exploiting the Hugging Face inference endpoints \cite{HFendpoint}.

\textbf{CodeLlama \cite{codellama}.} CodeLlama is a family of LLMs for code built on top of Llama2. CodeLlama has been trained on a corpus of 500B tokens featuring 85\% of code and 15\% of natural language. The models also underwent fine-tuning of the instruction to answer questions and perform required tasks. We adopt CodeLlama 7B, 13B, and 34B. In this case, the Hugging Face inference endpoints \cite{HFendpoint} have been used to generate predictions (\ie judgments).

\textbf{GPT-3.5-turbo and GPT-4-turbo \cite{chatgpt}.} Both models are behind OpenAI ChatGPT and have been shown to be able to handle code-related tasks \cite{tian2023chatgpt,le2024software,guo2024exploring}. Being the models on top of which a chatbot has been built, both GPT variants have been trained to follow instructions for required tasks. While the technical details behind these models are not publicly available, we can safely claim that (i) they are substantially larger than the DeepSeek Coder/CodeLlama variants previously described, with rumors placing GPT-3.5 and GPT-4 at over 150B and 1.5T trainable parameters, respectively; and (ii) their training set features massive amounts of textual data, including code mined from GitHub repositories. We exploit these models through the ChatGPT APIs.

\subsection{Context Selection: Evaluation Datasets}
\label{sub:datasets}
To assess the effectiveness of LLMs as a judge for \CG and \emph{summarization} we exploit two different datasets. 

\subsubsection{Code generation: CoderEval \cite{yu2024codereval}} Presented at ICSE'24, CoderEval is the most recent benchmark for code generation proposed in the literature. It features 460 code generation problems, 230 in Java and 230 in Python. Each problem is composed by (i) a natural language description specifying requirements for a function to implement; (ii) a target function showcasing a possible correct implementation; and (iii) a test suite to assess the correctness of automatically generated solutions. As detailed in \secref{sub:collection}, we take advantage of CoderEval to verify whether LLMs are able to judge the correctness of a given function. In particular, we will feed the LLM with a prompt that includes the problem description and the expected function signature. Then, we ask the model to judge if a \emph{candidate} implementation is correct or not. In addition, because the test suite is available, we can check whether the \emph{candidate} is actually correct and, as a consequence, whether the LLM's judgment was appropriate. For example, if the LLM says that \emph{candidate} correctly implements the requirements in the problem description, but the tests fail, we can detect a misjudgment.

Before adopting all 460 CoderEval's code generation problems in our study, we performed a ``quality assurance'' procedure aimed at verifying that the test results from such a dataset were reliable. Indeed, if the test results cannot be trusted due to a weak test suite, this would introduce a notable bias in our study. 
For example, a wrongly implemented function passing the tests and being correctly judged by an LLM as ``wrong'' would be unfairly accounted for as an LLM's misjudgment. For this reason, we performed the following quality checks on the CoderEval benchmark. First, we ensure that the target function associated with each code generation problem (\ie the function that demonstrates a possible correct implementation) actually passes the test. This was not the case for 20 Java and 39 Python problems that we decided to exclude. Second, we checked whether there were code generation problems for which an empty implementation (\ie a function that only has a signature and an empty body) would pass the tests. This occurred for another 9 Java problems raising concerns about the quality of the accompanying test suites and, as a consequence, resulting in their exclusion from our study. Finally, for the remaining 201 Java problems we further inspected all those for which the target signature had a non-\texttt{void} return type and implemented for each of them a dummy function just featuring a \texttt{return} statement appropriate for the expected return type. For example, for functions expected to return an object, the dummy function only featured in its body a ``\texttt{return null;}'' statement, while for those returning an \texttt{int} a ``\texttt{return 0;}'' was placed. We found that for an additional 17 problems, a dummy function was sufficient to make the test suite passes. We performed a similar check on the remaining 191 Python problems. However, since Python functions do not have an explicit return type, we always included in the dummy function to test a ``\texttt{return None}'' statement. One of these dummy functions managed to pass the tests and has been excluded. In the end, we could rely on 184 of the 230 Java and 190 of the 230 Python code generation problems featured in CoderEval.

\subsubsection{Code summarization}
\label{sub:data_cs} 
Our starting idea was to use the dataset by Roy \etal \cite{roy:fse2021} for the \CS study: This is a large-scale dataset featuring 6,253 code summary evaluations performed by 226 developers on 2,292 summaries (on average, 2.73 human evaluations each). Each participant was asked to assess the quality of Java methods' summaries in terms of conciseness, fluency, and content adequacy, with participants giving a score on a scale from 1 to 5 to each quality attribute (the higher the better). Unfortunately, by inspecting the dataset we found it to be suboptimal for our study. This was mostly due to the fact that the agreement between the developers who judged each summary was in general quite low (\eg we found a Krippendorff's $\alpha$ \cite{krippendorff2011computing} of $\sim$0.2 when looking at the judgements for content adequacy, which highlights a weak agreement). While some level of subjectivity in the judgements is expected, we thought that using such a human judgement as ``oracle'' to assess the LLMs judgement could be problematic. On top of this, the dataset features automatically-generated summaries judged by humans which were the output of quite old techniques, with the newest published in 2020 \cite{haque:2020}. These summaries are not representative of what modern LLMs are able to do (\eg some of them even featured  $<$\texttt{UNK}$>$ tokens) and thus do not represent an ideal target for our study, since we want to assess whether LLMs-as-a-judge could work for summaries automatically generated by modern LLMs. Finally, the dataset by Roy \etal \cite{roy:fse2021} only features Java methods' summaries, while our aim is to perform also the \CS study on both Java and Python.

For these reasons, we built (and make publicly available \cite{replication}) our own dataset that features human judgments of 1,163 summaries. To build the data set, we selected the top 100 Java and the top-100 Python functions in terms of the number of statements they feature from the CoderEval \cite{yu2024codereval} benchmark (\ie the one used in the \CG study). We decided to focus on the longest functions since those are the ones for which a good summary is likely to make a difference in terms of code comprehension, and thus assessing the quality of summaries for these functions may make more sense. Among these 200 functions, we found one Java and one Python functions that were a duplicate and were thus removed from the set. For each of the remaining 198 functions (99 per language), we have the associated code summary written by the original developer of the function. Furthermore, we asked five LLMs (\ie  CodeLlama 7B, 13B, and 34B \cite{codellama}, GPT-3.5-turbo and GPT-4-turbo \cite{chatgpt}) to generate a summary for each of these 198 functions. This process should have resulted in 198 (manually written) + 198$\times$5 (automatically generated) = 1,188 summaries (594 per language). However, for a few Python functions some of the LLMs outputted an empty summary, leading to a total of 594 (Java) + 569 (Python) summaries. Note that we did not use LLMs belonging to the DeepSeek Coder family in building this dataset since, as explained later, we did not manage to make them work for the \CS judging task and we wanted to have the same LLMs both generating and judging summaries, to also study any form of self-bias (\eg the LLM judges the summary it generates better as compared to the summaries generated by other LLMs or written by humans). The prompt used to generate code summaries with LLMs is documented in our replication package \cite{replication}.

Once obtained the 1,163 summaries we split them among \developers human judges, making sure that each summary was assessed by three judges (for a total of 1,163 $\times$ 3 = 3,489 judgements). All \developers judges have a Master' degree in Informatics or Computer Science, four of them have a Ph.D. in Software Engineering. On average, they have 5.8 years of experience (min = 1, max = 17) in Java programming and 6.9 in Python programming (min = 4, max = 10).
Taking inspiration from the work of Roy \etal \cite{roy:fse2021}, we asked participants to evaluate the quality of the summary in three dimensions: \emph{content adequacy}, \emph{conciseness}, and \emph{fluency \& understandability}. Note that compared to Roy \etal, we changed \emph{fluency} into \emph{fluency \& understandability} since, when looking at code summaries generated by modern LLMs (or written by humans), it is unlikely to find nonfluent text. Instead, it is possible that the summary, while fluent in terms of used English, is difficult to understand, for example, for people not having a deep domain knowledge of the code. Each of the three ``quality attributes'' has been assessed on a scale from 1 to 5 (the higher the better). We give clear guidelines to participants on how to interpret the scores. For example, in the case of \emph{fluency \& understandability}, these were the indications provided:

\begin{etaremune}
\item The summary is easy to read and understand and does not require specific domain knowledge to be understood.
\item The summary is easy to read and understand, but may require some specific domain knowledge to be understood.
\item The summary is easy to read and understand for developers with experience with that system.
\item The summary is difficult to read and understand, but is grammatically correct.
\item The summary is difficult to read and understand and is grammatically incorrect.
\end{etaremune}

Complete guidelines are publicly available  \cite{replication}.

After having finalized the dataset, we again inspected the agreement among the humans judges for the three quality aspects using the Krippendorff $\alpha$ \cite{krippendorff2011computing}. The Krippendorff's $\alpha$ can be interpreted as follows: $<$ 0.10 = agreement equivalent to chance; 0.10–0.20 = weak agreement; 0.21–0.40 = fair agreement; 0.41–0.60 = moderate agreement; 0.61–0.80 = substantial agreement; 0.81–0.99 = near-perfect agreement; and 1 = perfect agreement. For the \emph{content adequacy} we obtained $\alpha$=0.81 for the Java instances and $\alpha$=0.69 for the Python instances, for \emph{conciseness} $\alpha$=0.58 (Java) and $\alpha$=0.57 (Python), and for \emph{fluency \& understandability} $\alpha$=0.62 (Java) and $\alpha$=0.56 (Python). Overall, the guidelines provided to the participants resulted in quite high agreement in all three quality attributes.

\subsection{Data Collection}
\label{sub:collection}

\subsubsection{Code Generation} CoderEval \cite{yu2024codereval} provides a set of code generation problems, tests to assess the accuracy of candidate solutions, and a target, exemplar, generation. The target implementations for 374 code generation problems we consider in our study (184 Java and 190 Pythjon) represent the first batch of solutions we ask the LLMs to judge. 

\smallskip

\textbf{Experimented prompts.} We experiment with four different prompts for the \CG judging task. The first, referred to as ``\emph{zero shot}'', is a prompt inspired by Weyssow \etal \cite{Weyssow:llm-as-judge-se} who defined prompts to assess the extent to which LLMs are able to judge if an implementation satisfies specific non-functional requirements. Note that the focus of the authors \cite{Weyssow:llm-as-judge-se} is not on evaluating the judging capabilities of LLMs (as we do), but rather on exploiting them to check if LLMs can address non-functional requirements (more in \secref{sec:related}). The main idea we inherit from this work is to prompt  models for the rationale behind the score:\smallskip

{\color{darkgray}
\small
\emph{You will be provided with the description (``Description'') and the signature (``Signature'') of a \{language\} function to implement. You will also see a candidate implementation (``Candidate''). Your role is to evaluate the correctness of the Candidate, providing as output a rating and a rationale. Rate the Candidate with either 0 to 1:}

\begin{itemize}
\item[0.] \emph{**Wrong Implementation**: The implementation does not correctly implement the described function.}
\item[1.] \emph{**Correct Implementation**: The implementation correctly implements the described function.}
\end{itemize}

\noindent \emph{\# Description: \{description\}}

\noindent \emph{\# Signature: \{signature\}}

\noindent \emph{\# Candidate: \{candidate\}}
}

\smallskip

The definition of this prompt was driven by a trial-and-error procedure in which we tested multiple variants of the prompt with the eight LLMs subject of our study to ensure that the LLMs understood the task and provided an output that would be meaningful in most cases. This was quite challenging considering the variety of LLMs we considered. We tested the judging capabilities of the LLMs using different variations of the prompt on simple tasks, under the assumption that if they were not working on toy examples it would be very unlikely that they would work in a real scenario. By toy examples, we refer to situations in which it was evident that the candidate did not implement the requirements in the description (\eg the description asked for a method to sum two numbers, but the candidate was multiplying two numbers). 

The second prompt, named ``\emph{zero shot W/O rationale}'' is just a simplified version of the previous prompt in which we do not ask the LLM to also generate a rationale for the provided score.

The third prompt, referred to as ``\emph{automated CoT}'', is the classic ``Automated Chain-of-Thought'' prompting \cite{kojima2022large} vastly used in the literature to elicit reasoning in LLMs. This prompting strategy takes advantage of two steps. In the first, a \emph{reasoning extraction} prompt is provided, which, in our case, is the following:

\smallskip
{\color{darkgray}

\small
\emph{You will be provided with the description (``Description'') and the signature (``Signature'') of a \{language\} function to implement. You will also see a candidate implementation (``Candidate'').}

\noindent \emph{\# Description: \{description\}}

\noindent \emph{\# Signature: \{signature\}}

\noindent \emph{\# Candidate: \{candidate\}}

\smallskip

\noindent \emph{\# Question: is the Candidate correct according to all the functional requirements of the Description? Answer choices: ``Yes'' or ``No''.}

\smallskip

\noindent \emph{\# Reasoning: Let's think step by step.}

}
\smallskip

In the second step, the output of the above prompt (\ie $<$\texttt{REASONING}$>$) is used to build the \emph{answer extraction} prompt, which is basically identical to the previous one but it includes the generated $<$\texttt{REASONING}$>$ (rather than asking for it) and, after it, ends with: \emph{Therefore, the answer (``Yes'' or ``No'') is:}. 

Finally, the fourth experimented prompt named ``\emph{slow thinking}'', is a variant of the CoT prompt explicitly proposed by Tong and Zhang \cite{tong-zhang-2024-codejudge} to assess the code correctness in their approach named CodeJudge (built on top of GPT-3.5, more in \secref{sec:related}). We do not report the full prompt here for the sake of brevity, but it can be found in \cite{tong-zhang-2024-codejudge} and in our replication package \cite{replication}, together with all other prompts used in our study.

\smallskip

\textbf{Judging functions.} On top of the target implementations, we use the same four prompts to ask the LLMs to judge candidate implementations automatically generated by them. The prompt used for the code generation was composed of (i) the description of the code generation task as provided in CoderEval, and (ii) the signature of the target function in CoderEval. The main challenge at this point was to automatically extract the implemented function from the output of each LLM. Indeed, we observed that LLMs rarely output just the required function (even if explicitly prompted to do so), but tend to accompany the generated method with textual explanations, alternative implementations, \etc For this reason, we developed an \emph{implementation extractor} which exploits the lizard code analyzer \cite{lizard} to identify the first outputted function and consider it as the candidate implementation generated by the LLM. This script is publicly available \cite{replication} and, based on our manual analysis described below, always succeeded in extracting the candidate function, when present. There were, indeed, some code generation tasks for which the LLMs did not produce any output (documented in the results discussion --- \secref{sec:results}).

We use the same prompts exploited for the judging of the target implementations also for the 1,221 Java and 1,091 Python candidate implementations generated by the LLMs. Note that each LLM, besides judging methods outputted by other LLMs, also judges its own solution. This allows us to investigate whether a bias exists when LLMs are used as a judge for \CG.

Once all judging tasks have been run, a cleaning process on the LLMs' output was needed to collect their judgment. Also in this case we observed variety in the output template used by the LLMs, with some being more verbose than others and providing additional unrequested information (\eg a summary of the judged implementation). Thus, we extracted the judgments using a combination of parsing scripts (for simple cases) and manual extraction. To be confident about the extracted judgments, all of them have been manually checked by two authors, even the automatically collected ones. Also in this case, some LLMs failed to provide a judgment for some instances (as documented in the results discussion).

The above-described process provided us with a list of 80,556 total judgments referring to both functions written by humans (10,954 judgments) and automatically generated by LLMs (69,602). We describe in \secref{sub:analysis} how we analyze these data to answer our RQ.

\subsubsection{Code summarization} Our dataset features code summaries written by humans and generated by LLMs, which we ask LLMs to judge. 

\smallskip

\textbf{Experimented prompts.} Also for \CS we tested four different judging prompts. The fist, named \emph{zero shot}, is shown in the following:

{\color{darkgray}
\small
\emph{You will be provided with a \{language\} function (``Function'') and a textual summary of it (``Comment''). The goal of the Comment is to document the functionality implemented in the Function. Your role is to evaluate the Comment across three criteria, providing as output for each of them a rating (\# Rating) and a rationale (\# Rationale) as described in the following.} 

\smallskip

\noindent \emph{\# Evaluation Criteria}

\begin{itemize}
\item[*]\emph{Content adequacy: the extent to which the comment summarizes all information that can be inferred from the source code.}

\item[*]\emph{Conciseness: the extent to which the comment contains unnecessary information.}

\item[*]\emph{Fluency \& Understandability: the extent to which the comment is easy to read and understand.}
\end{itemize}

\noindent \emph{For each criterion, provide a score on a scale from 1 to 5: 1 (Very poor), 2 (Poor), 3 (Fair), 4 (Good), 5 (Very good).}\medskip

\noindent \emph{\# Function: \{function\}}

\noindent \emph{\# Comment: \{comment\}}

}

\smallskip

The second (\emph{zero shot + instructions}) is a more complex version of the \emph{zero shot} prompt in which we provided the LLMs with the same ``judging instructions'' we gave to the human judges when building our dataset (see \secref{sub:data_cs}). This basically means that we instruct the LLM on cases in which, for example, a 5 score for content adequacy was appropriate. Due to its length, we point the interested reader to our replication package for the full version of this prompt.

The third experimented prompt (\emph{automated CoT}) mirrors the one seen for \CG, with the first step featuring a \emph{reasoning extraction} prompt ending with ``\emph{\# Reasoning: Let us think step by step.}''. The output of this step is then used in the second step (\emph{answer extraction}) to produce the actual judgement. Finally, the fourth prompt (\emph{automated CoT + instructions}) is basically the \emph{automated CoT} prompt augmented with ``judging instructions'', as explained for \emph{zero shot + instructions}.

\smallskip

\textbf{Judging summaries.} Also in this case the extraction of the judgments required some manual effort due to models outputting unrequested information and not adopting a consistent output format when reporting their judgments. We developed extraction scripts and verified each piece of collected information manually. Cases where the script did not report any information were also inspected manually, to make sure no judgment was left out. We found out that all models belonging to the DeepSeek Coder family were not able to understand the \CS judging task, despite our efforts in trying to tune the prompt. Indeed, for over half of the summaries, the DeepSeek Coder models did not manage to output a judgment. This may be due to the multi-level judgment we require in this case (three criteria), which is substantially more complex than what was used for \CG. For the other models, we discuss the failure cases (\ie no judgment provided) in the results section.  

At the end, we obtained 22,304 total judgments referring to 3,865 summaries written by humans and 18,439 summaries automatically generated.

\subsection{Data Analysis}
\label{sub:analysis}
For both tasks we start by reporting the best-performing prompt, and we discuss in the paper only the results achieved with that prompt, while all others are included in our replication package \cite{replication}. It is important to give a clear definition of what we considered to be the best-performing prompt. Indeed, there are at least three dimensions to consider. First, different prompts may result in a different percentage of cases in which the LLMs fail to perform the judgment. These are cases in which the output of the LLMs, manually checked, did not feature the required judgment. Second, once collected the valid judgements, they may exhibit different levels of ``accuracy'' with different prompts. For example, a prompt may always succeed in outputting a judgement which, however, is always wrong (\eg correct functions always classified as wrong implementations), while another one may fail in a few cases but be quite accurate for the outputted judgements. Third, different LLMs may benefit more or less from different prompts. Since for both tasks there was one judge LLM which was the clear winner independently from the used prompt (\ie GPT-4-turbo), we selected as best-performing prompt the one ensuring the best performance on it (measured, as explained later, via inter-rater agreement metrics between the produced judgements and the oracles). Such a prompt also ensured a very low number of invalid outputs (\ie no judgement) outputted by GPT-4-turbo for both tasks and languages, not really posing a question about the selection of the best prompt to consider.

We report the percentage of cases in which each LLM did not manage to perform the judgment. We consider the judgment task failed if: (i) for the \CG task, the LLM did not output the required score; (ii) for the \CS task, the LLM did not output all three required scores.

\smallskip

\textbf{Code generation analyses.} We report eight confusion matrices (one per LLM) showing the percentage of (i) \emph{true positives}, \ie the candidate passes the tests and the LLM judges the candidate as a correct implementation of the requirements; (ii) \emph{true negatives}, \ie the candidate fails the tests and the LLM judges the candidate as not correct; (iii) \emph{false negatives}, \ie the candidate passes the tests but the LLM judges the candidate as not correct; and (iv) \emph{false positives}, \ie the candidate fails the tests but the LLM judges the candidate as correct. 

We complement the above analysis with the computation of an inter-rater agreement metric between the ``oracle'' (\ie test execution) and the LLMs' judgements. In particular, we report the Cohen's Kappa \cite{kappa} inter-rater agreement, which is suitable when only two categories of classifications are possible (in our case, the implementation is classified as either correct or wrong by both the LLMs and the tests execution). The interpretation is the same previously reported for the Krippendorff's $\alpha$.

We also verify whether the LLMs tend to give better score to code they generated as compared to code generated by others or written by humans. We run an unpaired Mann-Whitney test \cite{wilcoxon} comparing the scores assigned by each LLM to its own code to three other distributions, representing (i) the code generated by the other 7 LLMs; (ii) the code generated by the other LLMs excluding those belonging to the same family, with DeepSeek Coder, CodeLlama, and GPT being the three families; and (iii) the code written by humans. We use the unpaired Mann-Whitney test in this analysis because we are comparing distributions having a different number of elements. Indeed, let us assume we have 100 code generation tasks for which 5 LLMs (LLM$_1$, LLM$_2$, LLM$_3$, LLM$_4$, LLM$_5$) generated code. The same five LLMs are then asked to judge the generated code. To study the presence of self-bias, the scores assigned by LLM$_1$ to the 100 functions it generated are compared with the scores LLM$_1$ assigned to the 400 functions generated by LLM$_2$, LLM$_3$, LLM$_4$ and LLM$_5$. We account for multiple tests by adjusting the \emph{p}-values using the Benjamini-Hochberg procedure \cite{yoav:jstor1995}. We use Cliff's $d$ elta \cite{Cliff:2005} as effect size. Cliff's $d$ ranges in the interval $[-1,1]$ and is negligible for $|d| < 0.148$, small for $0.148 \le |d| < 0.33$, medium for $0.33 \le |d| < 0.474$, and large for $|d| \ge 0.474$.

Finally, we perform a manual analysis to disclose the reasons for the misjudgments made by the LLMs. We randomly selected for each LLM 15 cases of \emph{false positives} and 15 cases of \emph{false negatives}. We assigned 30 of the instances to inspect (15 \emph{false positives} and 15 \emph{false negatives}) to three authors, who independently defined one or more labels summarizing the reasons behind the LLM misjudgment. 
The definition of the labels was performed by looking at: (i) the prompt provided as input, (ii) the function to be judged, and (iii) the score and rationale provided by the LLM. For example, based on this information, we found that some \emph{false negatives} were due to the LLM assessing nonfunctional requirements rather than the code correctness, \eg giving a negative evaluation to a correct code because it was not efficient in terms of performance. After this first round, the three authors met and consolidated the labels used, involving then a fourth author to inspect all other remaining instances. In this step, each instance was assigned to two evaluators, who could reuse the already defined labels or add new ones when needed. The conflicts were solved by two additional authors not involved in the original labeling.

\begin{table*}[ht!]
    \centering
    \caption{Number (\#) and percentage (\%) of instances for which the LLMs did not manage to output a valid judgment.}
    \begin{tabular}{lrrcrrc|crrcrr}
    	 \toprule
         & \multicolumn{5}{c}{\textbf{Code generation}} &&& \multicolumn{5}{c}{\textbf{Code summarization}}\\\cline{2-6} \cline{9-13}
         \multirow{2}{*}{\textbf{LLM}} & \multicolumn{2}{c}{\textbf{Java}} && \multicolumn{2}{c}{\textbf{Python}} &&& \multicolumn{2}{c}{\textbf{Java}} && \multicolumn{2}{c}{\textbf{Python}}\\\cline{2-3} \cline{5-6} \cline{9-10} \cline{12-13}
         & \textbf{\#} & \textbf{\%} & & \textbf{\#} & \textbf{\%} &&& \textbf{\#} & \textbf{\%} && \textbf{\#} & \textbf{\%} \\
         \midrule
         DeepSeek Coder 1.3B & 20 & 1.42\% && 73 & 5.70\% &&& - & - && - & -\\
         DeepSeek Coder 6.7B & 145 & 10.32\% && 282 & 22.01\% &&& - & - && - & -\\
         DeepSeek Coder 33B & 156 & 11.10\% && 198 & 15.46\% &&& - & - && - & -\\
         CodeLlama 7B & 91 & 6.48\% && 139 & 10.85\% &&& 1 & 0.17\% && 1 & 0.18\%\\
         CodeLlama 13B & 184 & 13.10\% && 211 & 16.47\% &&& 23 & 3.87\% && 1 & 0.18\%\\
         CodeLlama 34B & 145 & 10.32\% && 132 & 10.30\% &&& 2 & 0.34\% && 20 & 3.51\%\\
         GPT-3.5-turbo & 4 & 0.28\% && 3 & 0.23\% &&& 0 & 0.00\% && 0 & 0.00\%\\
         GPT-4-turbo & 0 & 0.00\% && 7 & 0.55\% &&& 0 & 0.00\% && 0 & 0.00\%\\
         \bottomrule
    \end{tabular}
    \label{tab:failureJudgment}
\end{table*}

\smallskip

\textbf{Code summarization analyses.} We show three scatterplots for each judging LLM, one for each of the three criteria used in the assessment of summary quality (\ie content adequacy, conciseness, fluency \& understandability). The scatterplots allow us to visually identify the existence of a relationship between the quality assessments made by humans and by each LLM. 

As an agreement metric, we compute the Krippendorff's $\alpha$ \cite{krippendorff2011computing} for each of the three assessed quality attributes. Since we have three human judgements for each summary, we use their median score as the ``oracle'' against which the LLM judgements are compared. Given the rather substantial agreement we observed in the human judgements, this choice is not expected to substantially influence the achieved results.

\begin{table*}[ht!]
\centering
\caption{Code Generation: Kappa score between the LLMs' judgements (correct/wrong) and the output of test execution (pass/fail). Values below 0.10 indicate the complete lack of agreement.}
\scriptsize
\begin{tabular}{l|rrrrrrrr}
\toprule
\multirow{2}{*} &{\bf DSC 1.3B} & {\bf DSC 6.7B} & {\bf DSC 33B} & {\bf CL 7B} & {\bf CL 13B} & {\bf CL 34B} & {\bf GPT-3.5} & {\bf GPT-4} \\

\bottomrule\midrule
\multicolumn{9}{c}{\bf Java} \\ \midrule\toprule
\bf zero-shot & -0.15 & 0.00 & 0.14 & -0.01 & 0.13 & 0.16 & 0.10 & \bf 0.17 \\
\bf zero-shot W/O rationale & 0.05 & 0.00 & 0.11 & -0.06 & 0.06 & \bf \underbar{0.21} & 0.07 & 0.16 \\
\bf automated CoT & 0.03 & 0.03 & 0.15 & 0.04 & 0.05 & 0.10 & 0.16 & \bf \underbar{0.21} \\
\bf slow-thinking & 0.02 & 0.03 & 0.06 & 0.00 & 0.04 & 0.06 & 0.07 & \bf 0.15 \\
\bottomrule\midrule
\multicolumn{9}{c}{\bf Python} \\ \midrule\toprule
\bf zero-shot & 0.02 & -0.01 & \bf 0.10 & -0.02 & -0.01 & 0.04 & 0.05 & 0.07 \\
\bf zero-shot W/O rationale & 0.01 & 0.01 & \bf 0.10 & -0.04 & 0.02 & 0.03 & 0.03 & 0.05 \\
\bf automated CoT & -0.04 & 0.03 & 0.05 & 0.01 & 0.05 & 0.03 & 0.09 & \bf 0.10 \\
\bf slow-thinking & 0.00 & 0.08 & 0.06 & 0.00 & 0.01 & 0.06 & 0.06 & \bf \underbar{0.11} \\

\bottomrule
\end{tabular}
\label{tab:kappaRQ1}
\end{table*}

Also, for the \CS judgments, we verify whether the LLMs tend to be biased towards the summary they generate. This is done with a statistical analysis that mirrors that described for \CG.

Although we also considered performing the qualitative analysis of the reasons behind the misjudgments for \CS (similarly to what was done for \CG), we found it difficult to run it systematically, since we are dealing with three evaluation criteria and a 5-point score for each criteria, thus blurring the definition of false positives/negatives. 

\newcommand{\rulec}{\arrayrulecolor{black}\specialrule{0.1em}{\abovetopsep}{0pt}}%
\definecolor{rowsmallest}{RGB}{239, 239, 239}
\definecolor{rowlargest}{RGB}{190, 190, 190}
\section{Results Discussion} \label{sec:results}

\subsection{Code Generation} \label{sec:results_cg}

\begin{figure*}[htbp]
    \centering
    \subfloat[Java\label{fig:bool_judge_java}]{
        \includegraphics[width=0.48\textwidth]{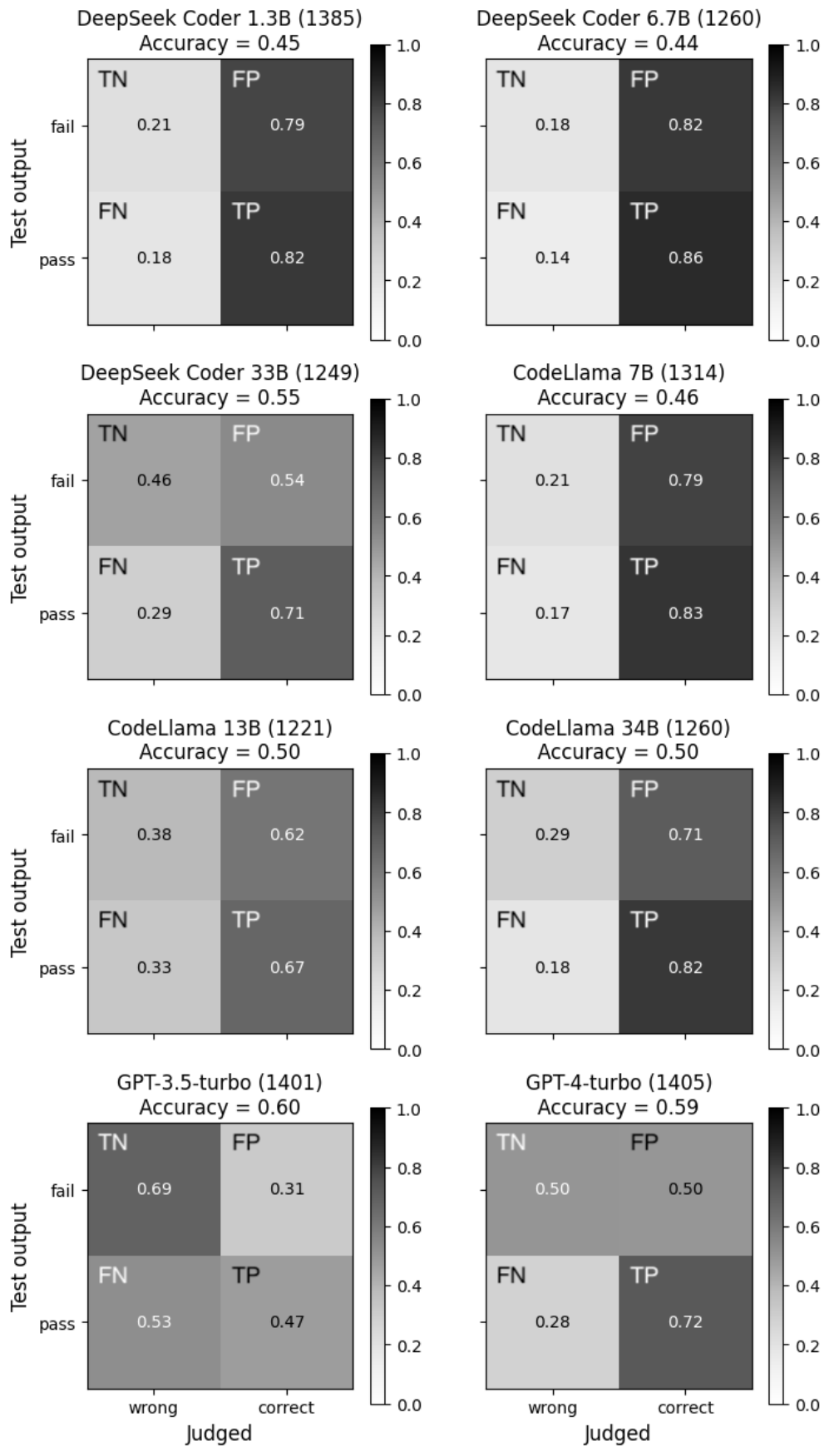}
    }
    \hfill
    \subfloat[Python\label{fig:bool_judge_python}]{
        \includegraphics[width=0.48\textwidth]{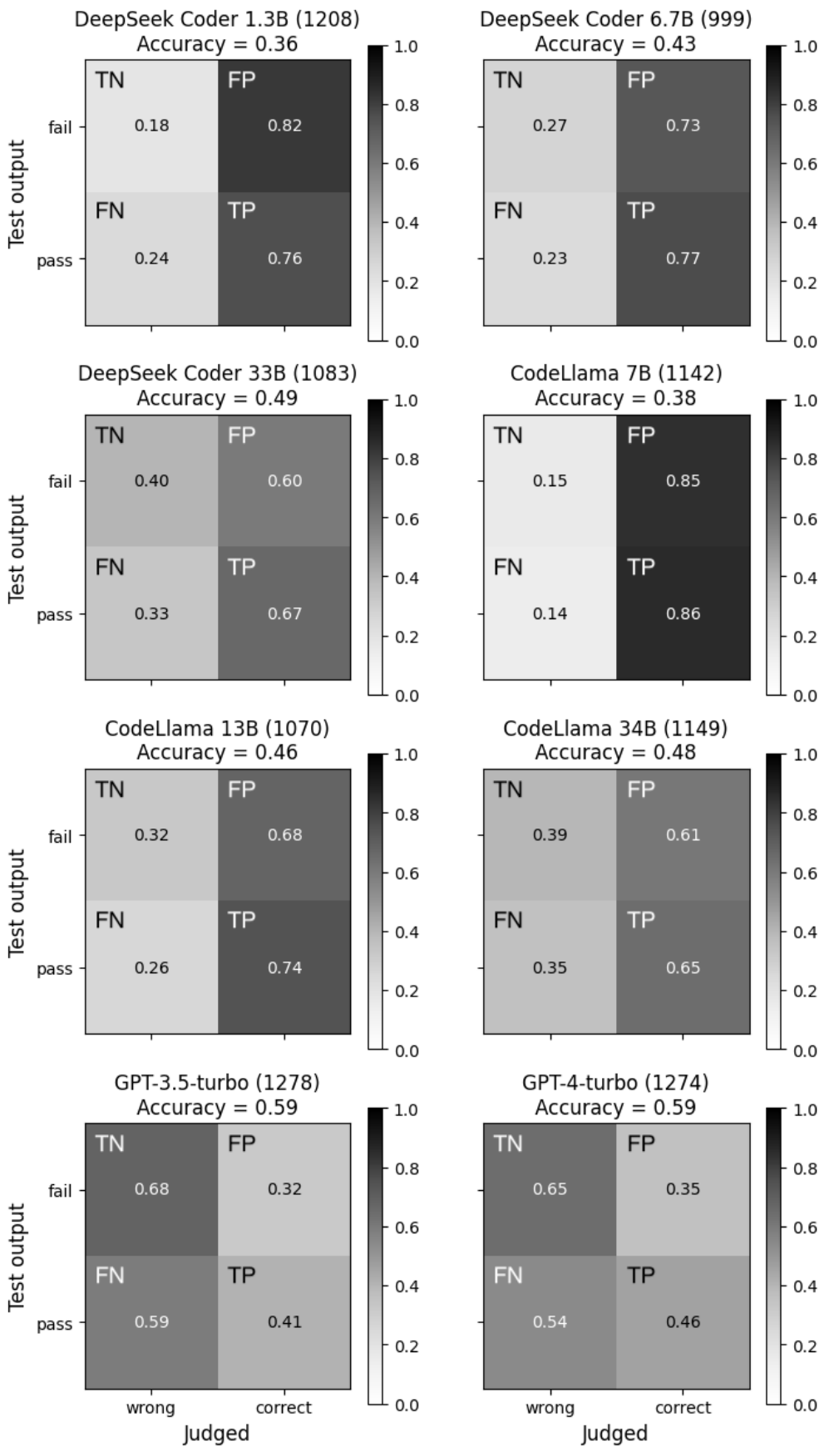}
    }
    \caption{Code Generation: Confusion matrices for LLM's judgment for Java (left) and Python (right). The $x$-axis reports the given judgment (\ie the function is a wrong or correct implementation of the textual requirements), while the $y$-axis shows the results of the test execution (\ie pass/fail). Within each matrix, true negatives are in the top-left box, true positives in the bottom-right.}
    \label{fig:bool_judge}
\end{figure*}

The left part of \tabref{tab:failureJudgment} shows the number and percentage of times for which using the best-performing prompt (\ie \emph{automated CoT}), each of the LLM-as-a-judge did not manage to output a judgment for the \CG scenario. As it can be seen, larger LLMs are almost always able to understand the required task  and execute it. For example, GPT-4-turbo always outputs a valid correctness judgement when dealing with Java functions, and it only fails for 7 Python functions. Instead, smaller LLMs with parameters of a few billion tend to fail in $\sim$15\% of cases, with minor differences observed between different LLMs (see \tabref{tab:failureJudgment}).

Understanding a task to be executed does not imply a correct execution (\ie a meaningful judgment), and this leads to the focus of our study. \tabref{tab:kappaRQ1} shows the results of Cohen's Kappa between the judgments given by the 8 LLMs and the test results (pass/fail). In this case, we report the data for the four experimented prompts, to (i) justify our selection of the \emph{automated CoT} prompt as the best-in-class prompt to discuss in the paper; and (ii) showing that our findings are not strongly impacted by the choice of the prompt. In \tabref{tab:kappaRQ1} the higher the agreement, the better the judgment. Recall that anything below 0.10 indicates the lack of agreement between the LLMs' correctness judgements and the test results. To save space, \tabref{tab:kappaRQ1} uses acronyms for LLMs (DSC = DeepSeek Coder, CL = CodeLlama). 

The first observation that can be made is that our findings do not substantially change using different prompts. In fact, when focusing on the single LLM in isolation (columns in \tabref{tab:kappaRQ1}), it can be seen that the variations in Kappa scores are rather minor using different prompts. Second, when considering the overall performance achieved in both languages (Java in the upper part of \tabref{tab:kappaRQ1}, and Python in the lower part), GPT-4 is the best performing LLM. Still, Cohen's Kappa between the judgments of GPT-4 and the ground truth is 0.21 for Java (with 0.20 being the lower limit at which the agreement can be considered \emph{fair}) and 0.10 for Python (\ie the lower limit at which the agreement can be considered \emph{weak}). Smaller LLMs are simply unable to assess code correctness. Indeed, DeepSeek Coder 1.3B and 6.7B as well as CodeLlama 7B obtain very low (even negative) agreement values in both languages (see \tabref{tab:kappaRQ1}). Larger models (DeepSeek Coder 33B, CodeLlama 34B, and GPT-3.5) tend to perform better, with performance closer to that of GPT-4.

To better understand what these agreement coefficients indicate, \figref{fig:bool_judge} shows the confusion matrices that depict the boolean judgments given by the eight LLMs to the correctness of the evaluated Java (left) and Python (right) functions. These refer to the judgments produced with the \emph{automated CoT} prompt.

The $x$-axis reports the given judgment (\ie the function is a wrong or correct implementation of the textual requirements), while the $y$-axis shows the results of the test execution (\ie pass/fail). To better understand the content of these matrices, let us discuss the findings for DeepSeek Coder 1.3B on the Java dataset (top-left confusion matrix). The top left cell reports the percentage (21\%) of failing implementations (\ie wrong implementations according to the test execution) which are correctly judged by the LLM as wrong (\emph{true negatives}), while the top-right cell shows the percentage (79\%) of misjudgments in this category, namely \emph{false positives}, \ie failing implementations judged as correct. The bottom row of the matrix shows instead the LLM's judgments related to passing implementations, with 18\% of them misjudged as wrong (\emph{false negatives}) and 82\% properly judged as correct (\emph{true positives}). The darker the background color of a cell, the higher the percentage of judgments falling into that category. This means that a successful judge would have the top left and bottom right cells of its confusion matrix substantially darker than the top right and bottom left cells. The numbers in parentheses in addition to the name of each model indicate the total judgments that the LLM successfully outputted. Finally, below each model's name we also report the overall accuracy computed based on the confusion matrix.

\figref{fig:bool_judge} shows that being the best among the experimented models does not make GPT-4 a ready and reliable solution for the automatic judgement of code correctness. In fact, if we look at the confusion matrices of its judgments for Java, we see that GPT-4 correctly classifies 72\% of the correct implementations but misjudges 50\% of the wrong ones. Basically, GPT-4 has a tendency to positively judge implementation, missing code bugs, and / or lack of features explicitly requested in the input requirements text. In the Python data set, instead, GPT-4 correctly classifies 46\% of the correct implementations and misjudges 35\% of the wrong ones, showing an inverse trend compared to Java. In general, even the best-performing model (GPT-4) often misjudges the correctness of the code.
 
One may conjecture that a possible reason for the observed wrong judgments could be the fact that the LLM has a ``limited view'' of the coding context when judging. Indeed, the code generation problems in CoderEval feature both functions which are self-contained (\ie they only have dependencies towards the Java/Python standard library), as well as functions which exploit \eg utility functions from third-party libraries. The latter represents more challenging code generation and judging scenarios. Indeed, the LLM-as-a-judge may, for example, complain about the usage in the judged function of a call towards a ``non-existing'' (\ie not visible) function. To account for such a scenario, we repeated all our analyses when considering only the 80 Java and 58 Python functions in CoderEval that do not have external dependencies. Our findings did not change in terms of the effectiveness of judgment and the identification of the best judge (see the replication package \cite{replication}), dismissing the lack of coding context as a major factor influencing the judging abilities of LLMs.

\begin{table*}[ht!]
    \centering
    \scriptsize
    \caption{Code Generation (Java): Average of differences between the LLM judgments (0 or 1) and the ground truth (\ie 1 if the method passes the test and 0 otherwise). Last three columns report adj. $p$-value (Mann-Whitney test, p-value adjusted using the Benjamini-Hochberg procedure) and effect size when comparing the judgments each LLM gave to functions it generated against those it gave when judging functions  (i) generated by all other LLMs, (ii) generated by all other LLMs but those belonging to the same family, and (iii) written by humans.}\vspace{-0.3cm}
    \begin{tabular}{lrrrrrrrr|r|rrr}
    \toprule
     & {\bf DSC} & {\bf DSC} & {\bf DSC} & {\bf CL} & {\bf CL} & {\bf CL} & {\bf GPT} & {\bf GPT} & {\bf Human} & {\bf Own \emph{vs}} & {\bf Own \emph{vs}} & {\bf Own \emph{vs}}\\
    & {\bf 1.3B} & {\bf 6.7B} & {\bf 33B} & {\bf 7B} & {\bf 13B} & {\bf 34B} & {\bf 3.5} & {\bf 4} & {\bf Written} & {\bf LLMs} & {\bf LLMs $\setminus$ F} & {\bf Human}\\
    \midrule
    
    {\bf DSC 1.3B} & \textcolor{gray}{0.58} & \textcolor{gray}{0.57} & \textcolor{gray}{0.43} & \textcolor{gray}{0.60} & \textcolor{gray}{0.43} & \textcolor{gray}{0.49} & \textcolor{gray}{0.41} & \textcolor{gray}{0.47} & \textcolor{gray}{-0.25} & \textcolor{gray}{ (N)} & \textcolor{gray}{ (N)}  & \textcolor{gray}{*** (L)}\\
    {\bf DSC 6.7B} & \textcolor{gray}{0.55} & \textcolor{gray}{0.49} & \textcolor{gray}{0.56} & \textcolor{gray}{0.57} & \textcolor{gray}{0.47} & \textcolor{gray}{0.51} & \textcolor{gray}{0.56} & \textcolor{gray}{0.60} & \textcolor{gray}{-0.24} & \textcolor{gray}{ (N)}  & \textcolor{gray}{ (N)}  & \textcolor{gray}{*** (L)}\\
    {\bf DSC 33B} & 0.31 & 0.36 & 0.24 & 0.33 & 0.29 & 0.30 & 0.29 & 0.28 & -0.36 & (N) & (N) & *** (L)\\
    {\bf CL 7B} & \textcolor{gray}{0.54} & \textcolor{gray}{0.55} & \textcolor{gray}{0.48} & \textcolor{gray}{0.59} & \textcolor{gray}{0.49} & \textcolor{gray}{0.49} & \textcolor{gray}{0.42} & \textcolor{gray}{0.48} & \textcolor{gray}{-0.22} & \textcolor{gray}{ (N)}  & \textcolor{gray}{ (N)}  & \textcolor{gray}{*** (L)}\\
    {\bf CL 13B} & 0.30 & 0.33 & 0.29 & 0.35 & 0.26 & 0.33 & 0.43 & 0.34 & -0.42 & (N) & (N) & *** (L)\\
    {\bf CL 34B} & 0.53 & 0.47 & 0.41 & 0.48 & 0.41 & 0.37 & 0.46 & 0.42 & -0.27 & (N) & (N) & *** (L)\\
    {\bf GPT-3.5} & 0.08 & 0.12 & 0.03 & 0.14 & 0.05 & 0.00 & 0.13 & 0.13 & -0.72 & (N) & (N) & *** (L)\\
    {\bf GPT-4} & 0.19 & 0.27 & 0.24 & 0.27 & 0.30 & 0.21 & 0.35 & 0.41 & -0.47 & ** (N) & ** (N) & *** (L)\\\midrule
    \textbf{\textcolor{gray}{Average (all)}} & \textbf{\textcolor{gray}{0.28}} & \textbf{\textcolor{gray}{0.31}} & \textbf{\textcolor{gray}{0.24}} & \textbf{\textcolor{gray}{0.32}} & \textbf{\textcolor{gray}{0.26}} & \textbf{\textcolor{gray}{0.24}} & \textbf{\textcolor{gray}{0.33}} & \textbf{\textcolor{gray}{0.32}} & \textbf{\textcolor{gray}{-0.45}} & \textbf{\textcolor{gray}{-}}  & \textbf{\textcolor{gray}{-}}  & \textbf{\textcolor{gray}{-}}\\
    \textbf{Average (large)} & \textbf{0.39} & \textbf{0.41} & \textbf{0.32} & \textbf{0.42} & \textbf{0.34} & \textbf{0.34} & \textbf{0.38} & \textbf{0.39} & \textbf{-0.37} & \textbf{-} & \textbf{-} & \textbf{-}\\
    
    \bottomrule
    \multicolumn{13}{c}{Adjusted $p$-values: * $<$0.05, ** $<$0.01, *** $<$0.001. Cliff delta: N=Negligible, S=Small, M=Medium, L=Large}
    \end{tabular}
    \label{tab:bias_cg}
    \end{table*}   
    
\subsubsection{Self-bias Analysis}

We also analyzed whether LLMs are biased towards code they generate, providing better judgment for it. \tabref{tab:bias_cg} shows a ``\emph{bias coefficient}'' of the LLMs-as-a-judge (rows) with respect to the code generated by the generator LLMs or manually written by humans (columns). We report the data for the Java dataset, since the findings for Python (available in \cite{replication}) are aligned. The coefficient, which ranges in $[-1, 1]$, is the average difference between the judgments (either 0 or 1) given by the LLM-as-a-judge and the ground truth (\ie 1 if the implementation passes the test suite and 0 otherwise). Given an LLM-as-a-judge ($LLM_J$), and a specific set of $I_G$ implementations to judge (\eg all methods generated by a specific generator $G$), the  \emph{bias coefficient} ($bc$) is computed as follows:
$$
bc_{J,G} = \frac{\sum_{i\in I_G} J_i - \sum_{i\in I_G} O_i}{|I_G|}
$$

\noindent where $J_i$ is the judgement (0 or 1 in this scenario) given by the judging LLM ($LLM_J$) to the $i^{th}$ code implementation to judge, while $O_i$ is the oracle score for that same code implementation (again, 0 or 1, depending on the tests outcome). Basically, if $bc_{LLM_J}$ is positive, this means that $LLM_J$ tends to overestimate the correctness of the assessed implementations (positive bias), while if $bc_{LLM_J}$ is negative it tends to have a negative bias. Note that a $bc_{LLM_J} = 0$, while indicating the absence of bias, does not imply the correctness of the judgements. For example, let us assume that $LLM_J$ has $n = 10$ implementations to judge, with the first five passing the test execution and the last five failing it. Let us also assume that $LLM_J$ misjudges all 10 instances, reporting the first five as wrong and the last five as correct. Both $\sum_{i=1}^{n} J_i$ and $\sum_{i=1}^{n} O_i$ will be equal 5, so their difference will be 0, indicating the lack of bias but not the correctness of the judgements.

We report in grey the assessments made by the LLMs which our former analyses revealed to be completely off in the judgments (DeepSeek Coder 1.3B, DeepSeek Coder 6.7B, CodeLlama 7B), since any conclusion  based on these judgments would be irrelevant. The last three columns of \tabref{tab:bias_cg} show the results of the Mann-Whitney tests (adj. $p$-value and effect size) in which we compare the judgments each LLM gave to functions it generated against those it gave when judging functions (i) generated by all other LLMs, (ii) generated by all other LLMs but those belonging to the same family (LLM $\setminus$ F), and (iii) written by humans. 

From \tabref{tab:bias_cg} two main observations can be made. First, with the exception of GPT-4, the LLMs are not biased in the judgements. If we focus on the five most reliable judges, GPT-3.5, DeepSeek Coder 33B, CodeLlama 13B, and CodeLlama 34B do not overestimate the quality of the code they generated more than they do with the code generated by other LLMs. For example, GPT-3.5 adds on average 0.13 points to the code it generated, and 0.14 points to the code generated by CodeLlama 7B. On the other hand, GPT-4 tends to slightly overestimate its own code more as compared to the code generated by other LLMs, revealing some sort of self-bias. Still, the statistical test shows that such a self-bias is minimum (\ie significant $p$-value but accompanied by a negligible effect size). The last two rows of \tabref{tab:bias_cg} show the average bias coefficient exhibited by the LLMs-as-a-judge for each code generator, either an LLM or a human. In grey we report the average across all eight judging LLMs, while in black when only considering the five that performed the best. In general, all LLMs tend to overestimate the correctness of the code generated by all other LLMs.

Second, it is interesting to see that LLMs have a higher chance of overestimating the quality of the code generated by LLMs rather than the code written by humans. Such a result can be seen from (i) the average bias coefficient achieved by the human written code (-0.37), indicating that the correctness of the code written by humans is systematically underestimated by LLMs, as compared to the coefficients of the LLMs which are all positive, and (ii) the results of the statistical tests (last column in \tabref{tab:bias_cg}). Indeed, all the judge models  underestimate the correctness of the human-written code in a statistically significant manner with a large effect size. This negative bias may be explained by the fact that human-written code is, from the LLMs' point-of-view, less natural as compared to code written by themselves or by other LLMs.

\subsubsection{Analysis of False Positives and Negatives}

We then looked at the reasons behind the \emph{false positive} and \emph{false negative} judgments performed by the eight models as identified in our manual analysis. We report in the replication package \cite{replication} the most frequent reasons per LLM, while we discuss here the general trend, considering the Java and Python instances as a single dataset. For false positives, most of the misjudgments (37\%) were due to \emph{uncaught wrong behavior}, with the LLM not catching bugs in the function (\eg the lack of checking for \texttt{null} values, explicitly asked in a code description). The second-most frequent reason (32\%) relates to the \emph{coding context}, and encompasses cases in which the LLM misjudges the function as correct since it is not aware of the need for using specific code elements (\eg objects) to successfully implement the requirements (and pass the tests). Note that while our former analysis showed that LLMs also frequently  fail in correctly judging self-contained functions, this does not exclude that in some cases the lack of context may result in a false positive classification. These cases may be addressed with more specific requirements text which, however, are not always available in code generation datasets. Indeed, we found that in 27\% of cases the judgment failure can be attributed to the \emph{ambiguous requirements text} used as input for the code generation. Those are not really failures of the judging LLMs, but are still representative of what would happen by employing LLMs as judges on code generation datasets.

When looking at false negatives, \emph{artificial hallucination} was the most popular reason (33\%) behind the misjudgments, with LLMs commenting about wrongly implemented statements which do not even appear in the judged function or commenting about non-implemented requirements which, instead, were implemented. The second top-reason for false negatives was the \emph{misunderstanding of code statements} (19\%), with LLMs just misunderstanding code and, as a consequence, wrongly assessing its correctness. Other less popular causes for false negatives included the limited \emph{coding context} (\ie the LLM complaining about the usage of code elements which were not visible in the code but, \eg part of the standard Java library), and \emph{focus on non-functional requirements} (\eg the LLM gives a negative judgment of the code not because of its correctness). Finally, a few unreliable test outcomes led to false negatives, \ie the implementation was actually wrong (so the LLM was right), but it was passing the CoderEval tests.

Finally, we also tried to understand to what extent the cases in which the LLMs managed to correctly judge code correctness were due to chance or to the actual LLMs' ability to understand the code. To do this, we collected the true positive instances outputted by the two best-performing LLMs, namely GPT-4 and GPT-3.5. The true positives are correct implementations that have been correctly judged as such by those models. To reduce the number of implementations to consider in this analysis (which, as it will be clear later, requires substantial manual work), we only considered correct instances properly judged by both models. This still resulted in 236 Java and 96 Python implementations. Then, we used the Universalmutator tool \cite{universalmutator} to generate, for each of the 332 implementations, a buggy version of them. Universalmutator is a regexp-based tool for mutating code across several languages, including Java and Python. We randomly selected one of the generated mutants for each of the 332 correct implementations, obtaining 332 pairs of $<$$correct\_code$, $mutated\_code$$>$. Finally, we add to each of these pairs a transformed version of $correct\_code$, being slightly different but semantically equivalent to it (thus, being correct). To obtain such a semantically-equivalent version, we applied a subset of the transformation operators implemented by Islam Rabin and Alipour in their ProgramTransformer tool \cite{Rabin:jss2022}. ProgramTransformer is meant to generate semantically-equivalent transformed programs, but it does not support Python (only Java and C\#). For this reason, we selected from its transformation operators those which can be easily applied to both Java and Python code, and we manually created from each pair $<$$correct\_code$, $mutated\_code$$>$ a triplet $<$$correct\_code$, $mutated\_code$, $sem\_eq\_code$$>$. The operators we considered are \cite{Rabin:jss2022}:

\begin{enumerate}
\item \emph{Loop Exchange}: It refactors a for {\tt loop} in an equivalent {\tt while} loop or vice versa.
\item \emph{Reorder Condition}: It switches the order of the {\tt true} and {\tt false} parts of an if statement.
\item \emph{Permute Statement}: It swaps two independent statements, ensuring no change in the code behavior.
\item \emph{Boolean Exchange}: It negates a boolean variable and propagates the change as needed to not change the behavior.
\item \emph{Variable Renaming}: It renames a local variable as {\tt var}.
\item \emph{Dead/Useless Code Insertion}: It inserts an unused statement or a statement which does not affect in any way the code behavior at a random place (\eg an unused variable declaration).
\end{enumerate}

The order in which we list the above operators reflects the order in which we tried to apply them given an instance. Indeed, some operators can only be applied if the $correct\_code$ to transform features specific code constructs (\eg \emph{Loop Exchange} requires the existence of a {\tt for}/{\tt while} loop) and, thus, cannot always be implemented. We prioritized these operators, and rely on \emph{Dead/Useless Code Insertion} (which can always be implemented) as a last resort when none of the previous operators can be applied. \tabref{tab:appliedOperators} shows the number and percentage of instances in which we managed to apply each operator for both Java and Python implementations. Worth noting is that we did not manage to equally use all the transformation operators, since some of them were rarely applicable on the benchmark's functions (\eg the \emph{Boolean Exchange} operator requires functions defining local boolean variables, which was rarely the case). Also, there are major differences across the two languages, again due to the types of functions present in the benchmark.

\begin{table}[h!]
    \centering
    \caption{Transformation operators applied to generate semantically-equivalent implementations of the 236 Java and 96 Python correct implementations.}
    \begin{tabular}{lrr}
         \toprule
         \textbf{Operator} & \textbf{\#Java (\%)} & \textbf{\#Python (\%)}\\\midrule
         \emph{Loop Exchange} & 59 (25\%)& 3 (3\%)\\
         \emph{Reorder Condition} & 45 (19\%)& 23 (24\%)\\
         \emph{Permute Statement} & 45 (19\%)& 30 (31\%)\\
         \emph{Boolean Exchange} & 0 (0\%)& 2 (2\%)\\
         \emph{Variable Renaming} & 42 (18\%)& 23 (24\%)\\
         \emph{Dead/Useless Code Insertion} & 45 (19\%)& 15 (16\%)\\
         \bottomrule
    \end{tabular}
    \label{tab:appliedOperators}
\end{table}

\begin{figure}
    \centering
    \includegraphics[width=\linewidth]{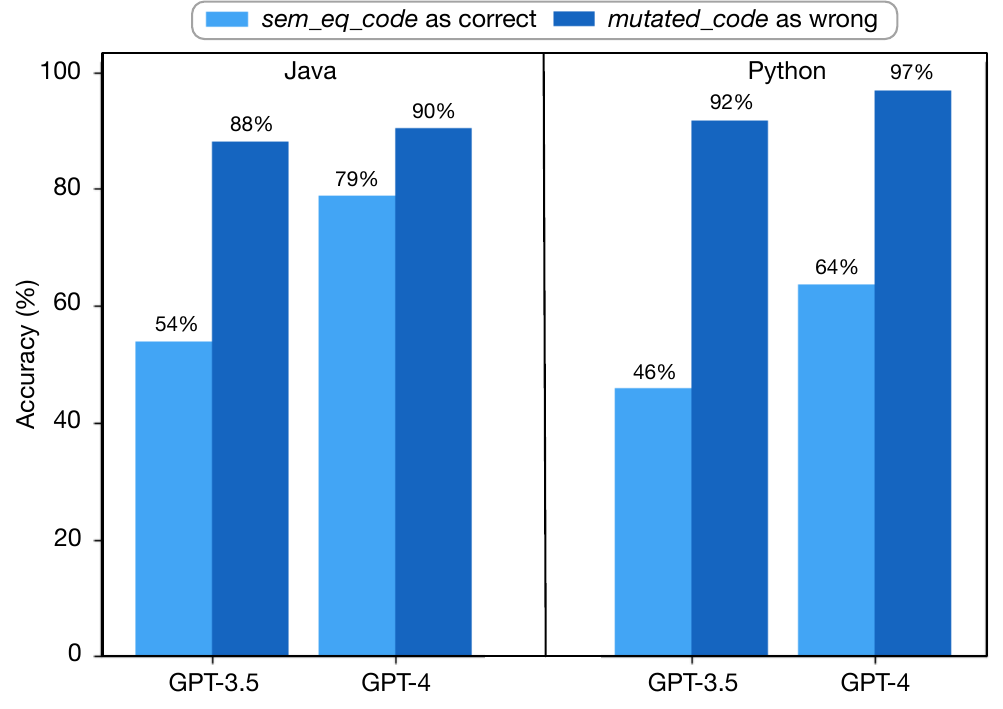}
    \caption{Code generation: Results of the mutants injection and semantically-equivalent code transformations applied on true positives.}
    \label{fig:mutation}
\end{figure}

Once obtained the triplets, for each of them we asked GPT-4 and GPT-3.5 to judge the correctness of the $mutated\_code$ and of the $sem\_eq\_code$. Since both LLMs already judged $correct\_code$ as correct, assuming that their correct assessment was due to understanding of the implemented code, in an ideal scenario they should (i) spot the bug in $mutated\_code$, classifying it as ``wrong'', and (ii) confirm the correctness of $sem\_eq\_code$.

\figref{fig:mutation} depicts the percentage of cases in which the two LLMs managed to (i) correctly classify $sem\_eq\_code$ as correct, thus confirming their original assessment of $correct\_code$ (light blue bars); and (ii) correctly classify $mutated\_code$ as wrong, thus spotting the bug we injected in $correct\_code$ (dark blue bars). The left part of \figref{fig:mutation} shows the results achieved in Java, while the right part refers to Python. The results in \figref{fig:mutation} paint a mixed picture. The models struggle to confirm their positive assessment of the correctness of $sem\_eq\_code$: GPT-3.5 reports only 54\% (Java) and 46\% (Python) of $sem\_eq\_code$ instances as correct. The situation is better for GPT-4, which reports 79\% (Java) and 64\% (Python) of $sem\_eq\_code$ as correct. Still, even the best-performing LLM changes its opinion about code correctness in 21\% (Java) and 36\% (Python) of cases, despite being presented with a semantically equivalent code.

On the positive side, the LLMs seem to be able to recognize buggy patterns related to the mutants we injected. In fact, both LLMs classified most of the $mutated\_code$ instances ($\geq$88\%) as wrong, with GPT-4 even achieving a 97\% in Python.

In summary, at least for GPT-4 the achieved findings allow to claim some code understanding capabilities of the LLMs. Still, this analysis also showed limitations that partially explain the mostly negative results we got when applying LLM-as a judge for \CG.

\subsection{Code Summarization}

We remind readers of the exclusion of the DeepSeek Coder family from the \CS study, as we did not manage to define a prompt to guide these LLMs towards an effective task resolution in most cases. We also point back to \tabref{tab:failureJudgment} which shows that the five LLMs considered in the \CS study rarely fail to produce a valid judgement (either correct or incorrect). As discussed later, the best prompt for this task was simple \emph{zero shot}. Thus, the data in \tabref{tab:failureJudgment} and the discussion of the results for \CS refer to this prompt (additional results available in \cite{replication}).

\begin{table*}[ht!]
        \centering
        \caption{Code Summarization: Krippendorff's $\alpha$ between the LLMs judgments and the median human rating. Values below 0.10 indicate the complete lack of agreement.}
        \scriptsize
        \begin{tabular}{l|rrrrr|rrrrr}
        \toprule
        
        & \multicolumn{5}{c|}{\bf Java} & \multicolumn{5}{c}{\bf Python}\\

        \bottomrule\midrule
        \multirow{2}{*} & {\bf CL 7B} & {\bf CL 13B} & {\bf CL 34B} & {\bf GPT-3.5} & {\bf GPT-4} & {\bf CL 7B} & {\bf CL 13B} & {\bf CL 34B} & {\bf GPT-3.5} & {\bf GPT-4}\\ \midrule\toprule
        & \multicolumn{10}{c}{\bf zero shot} \\ \midrule\midrule
        \it Content Adequacy & -0.04 & -0.02 & -0.11 & 0.18 & \bf 0.58 & -0.26 & -0.17 & -0.10 & 0.17 & \bf 0.63 \\
        \it Conciseness & -0.64 & -0.68 & -0.40 & -0.23 & \bf 0.40 & 0.06 & -0.23 & -0.10 & -0.37 & \bf 0.36\\
        \it Fluency \& Understandability & 0.04 & 0.16 & -0.02 & \bf 0.41 & 0.29 & 0.03 & -0.06 & 0.03 & 0.25 & \bf 0.44\\
        \midrule\toprule
        & \multicolumn{10}{c}{\bf zero shot + instructions} \\ \midrule\midrule
        \it Content Adequacy & -0.11 & 0.05 & -0.11 & 0.05 & \bf 0.52 & -0.42 & -0.14 & -0.19 & 0.07 & \bf 0.55 \\
        \it Conciseness & -0.26 & -0.82 & 0.02 & 0.13 & \bf 0.20 & -0.04 & -0.31 & -0.16 & \bf 0.29 & 0.03\\
        \it Fluency \& Understandability & -0.05 & -0.04 & 0.15 & \bf 0.25 & 0.23 & -0.18 & 0.04 & 0.03 & 0.21 & \bf 0.38\\
        \midrule\toprule
        & \multicolumn{10}{c}{\bf automated CoT} \\ \midrule\midrule
        \it Content Adequacy & 0.05 & 0.03 & 0.04 & -0.09 & \bf 0.60 & -0.07 & 0.01 & -0.02 & -0.19 & \bf 0.51 \\
        \it Conciseness & -0.06 & -0.42 & -0.10 & \bf 0.20 & 0.17 & 0.04 & -0.24 & 0.04 & 0.29 & \bf 0.31\\
        \it Fluency \& Understandability & 0.08 & -0.02 & 0.07 & 0.11 & \bf 0.22 & -0.14 & 0.00 & -0.03 & -0.07 & \bf 0.37\\
        \midrule\toprule
        & \multicolumn{10}{c}{\bf automated CoT + instructions} \\ \midrule\midrule
        \it Content Adequacy & -0.04 & -0.02 & 0.07 & 0.08 & \bf 0.58 & -0.19 & 0.06 & -0.02 & 0.04 & \bf 0.42 \\
        \it Conciseness & -0.06 & -0.43 & -0.22 & 0.11 & \bf 0.16 & -0.05 & -0.32 & -0.10 & 0.19 & \bf 0.42\\
        \it Fluency \& Understandability & 0.05 & -0.08 & 0.07 & 0.05 & \bf 0.41 & -0.03 & 0.04 & 0.08 & -0.06 & \bf 0.42\\

        \bottomrule
        \end{tabular}
        \label{tab:krippendorffRQ2}
        \end{table*}

\tabref{tab:krippendorffRQ2} reports the Krippendorff's $\alpha$ agreement score between the LLMs and human judgment for each quality criterion (\ie \emph{content adequacy}, \emph{conciseness}, and \emph{fluency \& understandability}) and when considering the four experimented prompts (\ie \emph{zero shot}, \emph{zero shot + instructions}, \emph{automated CoT}, \emph{automated CoT + instructions}).

When considering the agreement score obtained by the best-performing LLM (\ie GPT-4) across all three code summary quality criteria, the \emph{zero shot} prompt represents the best compromise. Indeed, while not ensuring the best judging performance for all quality criteria, it does not exhibit strong drops for a specific criterion as it happens with other prompts. For example, \emph{automated CoT} is the best performing in terms of \emph{content adequacy} on the Java data set (left part of \tabref{tab:krippendorffRQ2}), with $\alpha$ = 0.60 versus 0.58 of \emph{zero shot}. However, it suffers in judging \emph{conciseness} (0.17 versus 0.40 of \emph{zero shot}) and \emph{fluency \& understandability} (0.22 \emph{vs} 0.29). Similar observations can be made for the Python dataset (right part of \tabref{tab:krippendorffRQ2}).

Similarly to what observed for \CG, it is fair to claim that the choice of the prompt has some impact, but it does not substantially changes the overall findings: the Krippendorff's $\alpha$ of the LLMs look quite stable when changing prompt (with a few notable exceptions), and the best-performing LLMs is usually GPT-4 (again, with a few exceptions discussed in the following).

\begin{figure}
    \centering
    \includegraphics[width=\linewidth]{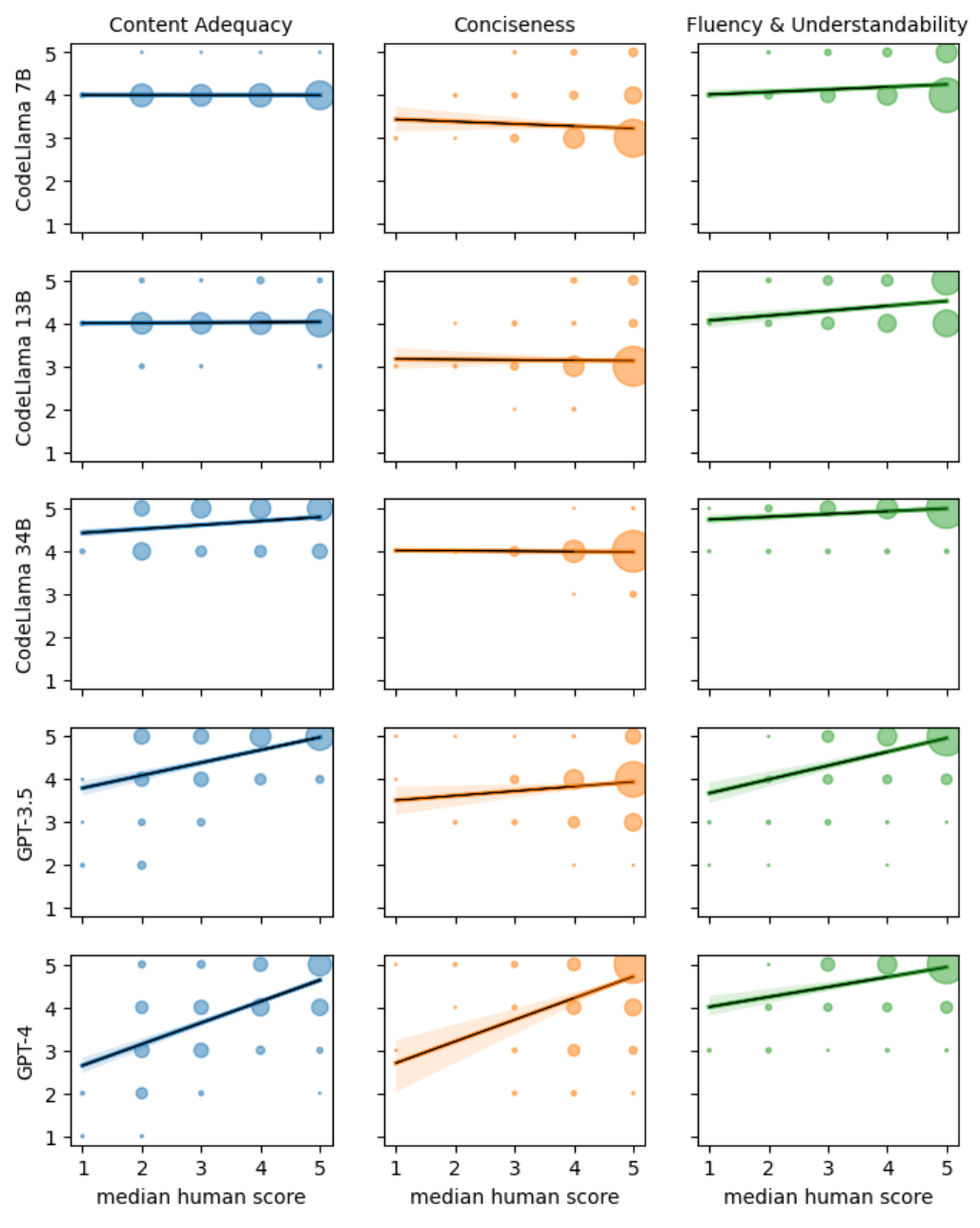}
    \caption{Code summarization (Java): Scatterplots relating human to LLM judgments. The three quality criteria subject of the judgment are shown as ``columns'', while the judging LLMs are ``rows''. Within each scatterplot the median of the ratings provided by the three human judges is on the $x$-axis, and the score assigned by the LLM is on the $y$-axis.}
    \label{fig:scatterplot}
\end{figure}

To better support the results discussion, \figref{fig:scatterplot} shows the scatterplots depicting, for each of the three quality criteria subject of the judgment (``columns'' in \figref{fig:scatterplot}) and for each LLM (``rows''), (i) the median of the ratings provided by the three human judges ($x$-axis), and (ii) the score assigned by the LLM ($y$-axis) when using the \emph{zero shot} prompt. For the sake of brevity, we show the graph only for the Java dataset, since the findings for Python are very similar (and available in \cite{replication}), as can also be seen from the agreement scores in \tabref{tab:krippendorffRQ2}. A blue dot placed at coordinates ($x$=2, $y$=3) in a graph of the first column indicates a summary for which the median human judgment for \emph{content adequacy} was 2, while the LLM score was 3. A successful judge would result in most of the points placed on the antidiagonal of the scatterplot (\ie the diagonal going from top right towards bottom left). The size of the dots in each graph is proportional to the number of cases that fall into the specific judging scenario. For example, if we focus on the graph related to CodeLlalma 7B judgments for \emph{conciseness} (first row, second column), the largest dot is placed in position ($x$=5, $y$=3) and features 354 cases (out of the 594 Java summaries), while the specular scenario ($x$=3, $y$=5) only features 2 cases.
To help interpret the scatter plot, we also include the regression line between the two variables.

\figref{fig:scatterplot} suggests the lack of a relationship between scores assigned by humans and those provided by the smaller LLMs used in our study (\ie CodeLlama 7B, 13B, 34B). Indeed, the regression line does not show any sort of trend aligned with the anti-diagonal, with almost a flatten line. The Krippendorff test (\tabref{tab:krippendorffRQ2}) confirms the lack of agreement between the judgments provided by the models in the CodeLlama family and those assigned by humans for all quality criteria and for both languages. In fact, most of the agreement scores are negative.

The situation is different when one looks at much larger models belonging to the GPT family. When judging \emph{content adequacy} and \emph{fluency \& understandability}, both models show the ability to mirror human judgment. Both GPT-3.5 and GPT-4 achieve positive Krippendorff alphas for \emph{content adequacy}: GPT-3.5 gets $\alpha=0.18$ on the Java and $\alpha=0.17$ on Python summaries; for GPT-4 these values go up to $\alpha=0.58$ (Java) and $\alpha=0.63$ (Python). The level of agreement achieved by GPT-4 is surprisingly high considering the challenges of a 5-level classification task and reveals the ability of GPT-4 to assess the quality of the information featured in the summary similarly to humans. GPT-4 is also the only LLM that reaches moderate agreement with humans when it comes to the \emph{conciseness} of the code summaries ($\alpha=0.40$ for Java and $\alpha=0.36$ for Python). Finally, GPT-4 also achieves a fair (Java $\alpha=0.29$) and a moderate (Python $\alpha=0.44$)  agreement with humans when evaluating \emph{fluency \& understandability}.

It is also interesting to note that the agreement between the best performing LLM (\ie GPT-4) and humans is the highest for the same quality aspects for which it is also the highest among humans. Indeed, as explained in \secref{sub:data_cs}, also humans had the highest agreement when judging the \emph{content adequacy of summaries} ($\alpha$=0.81 for Java and $\alpha$=0.69 for Python).

\begin{table*}[ht!]
    \centering
    \scriptsize
    \caption{Code Summarization (Java): Average of differences between the LLM judgments and the ground truth (\ie the median of the score given by the three human raters). Last three columns report adj. $p$-value and effect size when comparing the judgments each LLM gave to summaries it generated against those it gave when judging summaries  (i) generated by all other LLMs, (ii) generated by all other LLMs but those belonging to the same family, and (iii) written by humans.\vspace{-0.3cm}}
    \begin{tabular}{lrrrrr|r|rrr}
    \toprule
    
    & {\bf CL 7B} & {\bf CL 13B} & {\bf CL 34B} & {\bf GPT-3.5} & {\bf GPT-4} & {\bf Human Written} & {\bf Own \emph{vs} LLMs} & {\bf Own \emph{vs} LLMs $\setminus$ F} & {\bf Own \emph{vs} Human}\\
    \bottomrule\midrule
    
    \multicolumn{10}{c}{\bf Content Adequacy} \\ \midrule\toprule
    {\bf human} & 2.88 & 3.68 & 3.80 & 4.11 & 4.68 & 2.97 & - & - & - \\\midrule
    {\bf CL 7B} & \cellcolor{rowlargest}{1.14} & 0.33 & 0.21 & -0.11 & \cellcolor{rowsmallest}{-0.68} & 1.03 & *** (L) & *** (L) & (N) \\
    {\bf CL 13B} & \cellcolor{rowlargest}{1.17} & 0.36 & 0.29 & -0.14 & \cellcolor{rowsmallest}{-0.64} & 1.01 & (N) & *** (M) & *** (S) \\
    {\bf CL 34B} & \cellcolor{rowlargest}{1.65} & 1.04 & 0.71 & 0.81 & \cellcolor{rowsmallest}{0.32} & 1.43 & (N) & (N) & *** (S) \\
    {\bf GPT-3.5} & \cellcolor{rowlargest}{1.59} & 1.04 & 1.00 & 0.73 & \cellcolor{rowsmallest}{0.30} & 0.78 & (N) & *** (S) & (N) \\
    {\bf GPT-4} & \cellcolor{rowlargest}{0.63} & \cellcolor{rowsmallest}{0.13} & 0.20 & 0.39 & \cellcolor{rowsmallest}{0.13} & 0.29 & (N) & (N) & (N) \\
    \toprule\midrule
    
    \multicolumn{10}{c}{\bf Conciseness} \\ \midrule\toprule
    {\bf human} & 4.45 & 4.62 & 4.51 & 4.92 & 4.81 & 4.80 & - & - & - \\\midrule
    {\bf CL 7B} & \cellcolor{rowlargest}{-1.15} & -1.39 & -1.17 & \cellcolor{rowsmallest}{-1.71} & -1.67 & -1.53 & ** (S) & *** (S) & ** (S) \\
    {\bf CL 13B} & -1.37 & -1.55 & \cellcolor{rowlargest}{-1.34} & \cellcolor{rowsmallest}{-1.77} & -1.61 & -1.74 & (N) & (N) & * (S) \\
    {\bf CL 34B} & \cellcolor{rowlargest}{-0.49} & -0.63 & -0.50 & \cellcolor{rowsmallest}{-0.91} & -0.81 & -0.86 & *** (S) & *** (S) & *** (M) \\
    {\bf GPT-3.5} & -0.69 & \cellcolor{rowlargest}{-0.56} & -0.63 & -0.86 & -0.83 & \cellcolor{rowsmallest}{-1.10} & ** (S) & ** (S) & ** (S) \\
    {\bf GPT-4} & -0.08 & -0.25 & \cellcolor{rowsmallest}{-0.39} & \cellcolor{rowlargest}{0.06} & -0.08 & 0.00 & (N) & (N) & (N) \\
    \toprule\midrule

    \multicolumn{10}{c}{\bf Fluency \& Understandability} \\ \midrule\toprule
    {\bf human} & 4.40 & 4.68 & 4.81 & 4.67 & 4.82 & 3.71 & - & - & - \\\midrule
    {\bf CL 7B} & -0.21 & -0.51 & \cellcolor{rowsmallest}{-0.59} & -0.45 & -0.47 & \cellcolor{rowlargest}{0.52} & ** (S) & * (N) & *** (M) \\
    {\bf CL 13B} & 0.04 & -0.24 & \cellcolor{rowsmallest}{-0.42} & -0.05 & -0.36 & \cellcolor{rowlargest}{0.71} & (N) & (N) & *** (L) \\
    {\bf CL 34B} & 0.58 & 0.31 & \cellcolor{rowsmallest}{0.15} & 0.33 & 0.17 & \cellcolor{rowlargest}{1.16} & * (N) & (N) & *** (L) \\
    {\bf GPT-3.5} & 0.41 & 0.18 & \cellcolor{rowsmallest}{0.10} & 0.31 & 0.18 & \cellcolor{rowlargest}{0.56} & (N) & (N) & (S) \\
    {\bf GPT-4} & 0.46 & 0.16 & \cellcolor{rowsmallest}{-0.01} & 0.31 & 0.18 & \cellcolor{rowlargest}{0.76} & (N) & (N) & *** (M) \\
    
    \bottomrule
    \multicolumn{10}{c}{Adjusted $p$-values: * $<$0.05, ** $<$0.01, *** $<$0.001. Cliff delta: N=Negligible, S=Small, M=Medium, L=Large}
    \end{tabular}
    \label{tab:bias_cs}
\end{table*}

\subsubsection{Self-bias Analysis}

As in the case of \CG task, in this case we also assess the extent to which the LLMs are biased when judging the summaries they generate. \tabref{tab:bias_cs} shows the \emph{bias coefficient} for each LLM-as-a-judge (row) with respect to the summaries either automatically generated by the five subject LLMs or written by humans (column). The formula for computing the bias coefficient is the same as defined in \secref{sec:results_cg}, \ie the average of the differences between the LLM judgement and the ground truth (\eg the median of the score provided by human raters): It represents how much, on average, a summary created by a generator (column) is overestimated (positive values) or underestimated (negative values) by the judge (row). Unlike the \CG task, here the scores are between 1 and 5 for each of the three quality attributes (\ie \emph{content adequacy}, \emph{conciseness}, and \emph{fluency \& understandability}). In each row, we highlight in dark gray the strongest positive bias exhibited by the judging LLM and in light gray the strongest negative bias. For each quality attribute, we also report the average median score given by humans to the summaries generated by each LLM or written by humans (see the first row in each of the three subtables in \tabref{tab:bias_cs}). Finally, as already done for the \CG study, the last three columns of \tabref{tab:bias_cs} report the results of the statistical tests that compare the judgements each LLM gave to the summary it generates versus those (i) generated by other LLMs (\ie Own \emph{vs} LLMs), (ii) generated by LLMs of a different family (\ie Own \emph{vs} LLMs $\setminus$ F) and (iii) written by humans. Data in \tabref{tab:bias_cs} refer to the Java data set and, again, to the \emph{zero shot} prompting. Those related to Python and other prompts lead to the same conclusions and are thus included in our replication package \cite{replication}.

To better understand the content of the three subtables in \tabref{tab:bias_cs}, let us focus on \emph{content adequacy} (the top part of \tabref{tab:bias_cs}). From its first row (median human judgement), we can see that the participants evaluated the summaries generated by GPT-4 as the best overall in terms of information they feature (\emph{content adequacy}), with an average median score of 4.68. For comparison, human-written summaries were evaluated with an average median score of 2.97, which is considered worse than most LLM-generated ones. Although this is not the focus of our work (since it is not related to the capabilities of LLM as a judge), it is an interesting side finding that deserves further investigation (\ie Did LLM become better than humans at documenting code?). 

Moving to the bias coefficients, let us take the example of CodeLlalma 7B as a judge, which exhibits a bias coefficient of 1.14, which means that on average it overestimates the \emph{content adequacy} of the summaries it generates by 1.14 points over the median human judgement. This is the strongest positive bias it has (dark gray), while its strongest negative bias is towards summaries generated by GPT-4 (light gray).

\begin{table*}[ht!]
    \centering
    \caption{Content adequacy (Java): ranking of the generators of summaries according to each judge, including both humans and LLMs.\vspace{-0.2cm}}
    \scriptsize
    \begin{tabular}{l|rrrrr}
    \toprule
    {\bf Human} &  {\bf CL 7B} & {\bf CL 13B} & {\bf CL 34B} & {\bf GPT-3.5} & {\bf GPT-4} \\
    \midrule
    GPT-4 (4.68) & CL 7B (4.02) & GPT-4 (3.95) & GPT-4 (5.00) & GPT-4 (4.98) & GPT-4 (4.81)\\
    GPT-3.5 (4.11) & CL 13B (4.01) & human (3.92) & GPT-3.5 (4.92) & GPT-3.5 (4.84) & GPT-3.5 (4.51)\\
    CL 34B (3.80) & CL 34B (4.01) & CL 13B (3.90) & CL 13B (4.72) & CL 34B (4.80) & CL 34B (4.00)\\
    CL 13B (3.68) & GPT-3.5 (4.00) & CL 7B (3.86) & CL 7B (4.47) & CL 13B (4.72) & CL 13B (3.81)\\
    human (2.97) & GPT-4 (4.00) & GPT-3.5 (3.82) & CL 34B (4.45) & CL 7B (4.46) & CL 7B (3.51)\\
    CL 7B (2.88) & human (3.96) & CL 34B (3.75) & human (4.40) & human (3.75) & human (3.26)\\
    \bottomrule
    \end{tabular}
    \label{tab:cs_ca_ranking}
\end{table*}

To simplify the discussion of the possible bias exhibited by LLM judges, \tabref{tab:cs_ca_ranking} reports the classification of summary creators according to what concerns \emph{content adequacy} according to humans (first column) and each LLM judge (second to sixth columns). Similar tables are available in our replication package for the other two quality attributes \cite{replication} and for Python. 
 
Looking at \tabref{tab:bias_cs} for what concerns \emph{content adequacy}, it may seem that all models are positively biased toward CodeLlama 7B. However, the only two models exhibiting a statistically significant bias towards such a model are CodeLlama 7B itself and its 13B version. Also, looking at \figref{fig:scatterplot} and \tabref{tab:cs_ca_ranking}, it can be seen that these two models mostly act as ``constant'' classifiers when assessing \emph{content adequacy}, giving a score of 4 for almost every input. This results in a strong positive bias towards what are considered the worst summaries according to humans (\ie those generated by CodeLlama 7B). However, given the ``constant behavior'' of these models, we do not claim any sort of self-bias, but rather their (already discussed) lack of ability to act as judges for the quality of code summary.

Models from the GPT family tend to overestimate the \emph{content adequacy} of summaries, both those generated by automatic techniques and written by humans. However, their ranking of ``summaries generators'' (\tabref{tab:cs_ca_ranking}) is quite aligned to that of humans, with the only exception of considering as worst the summaries written by humans rather than those generated by CodeLlama 7B. Also, looking at these data, it is clear that GPT-4 is a major step ahead in judging capabilities as compared to GPT-3.5: In fact, the former only overestimates the \emph{content adequacy} of CodeLlama 7B by 0.63 points, on average, while the latter by 1.59 points. GPT-4 also does not exhibit any sort of self-bias, as confirmed by the statistical tests.

Moving to the summaries' \emph{conciseness}, most of the bias coefficients are negative for all LLMs, indicating that they tend to give lower scores as compared to humans. Also, excluding the judges mostly acting as constant classifiers (those from the CodeLlama family), again GPT-4 does not show any form of self-bias, while GPT-3.5 has a slight negative bias towards the \emph{conciseness} of summaries generated by the GPT family. Once again, no particular self-bias can be observed.

Finally, while all LLMs tend to overestimate \emph{fluency \& understandability} summaries written by humans, for this quality attribute, we also did not observe any form of self-bias.

\subsection{Discussion}
For both subject tasks (\ie \CG and \CS) we found that smaller LLMs (\eg DeepSeek Coder, CodeLlalma) struggle as judges, while the largest model we considered (\ie GPT-4-turbo) was the best in class. However, there are profound differences among the two tasks in what that means from a practical point of view. 

In the context of \CG, even the best-performing model (GPT-4) frequently misjudges the correctness of code, also showing different trends among the two experimented languages: When assessing Java code, GPT-4 lacked especially in the identification of wrong code implementations, while for Python it misjudged several correct implementations as wrong. Although some misjudgments can be addressed by providing LLMs with more detailed requirements about what is expected from the judged code, we also noted that even extremely large LLMs such as GPT-4 struggle in reasoning about code correctness. This would raise questions about the practical application of LLMs as a judge in \eg automated code review or bug fixing.  

The situation is quite different when it comes to the \CS task. Here, GPT-4 showed quite high agreement with human judgment of code summaries, especially when assessing \emph{content adequacy}. The higher effectiveness in assessing the quality of natural language text (rather than code correctness) may somehow be expected considering the vast amount of text seen by the LLM during training. 

Given the low correlation between human judgment of the quality of summaries and metrics frequently used to assess the performance of code summarization techniques (\eg BLEU \cite{papineni2002bleu}, ROUGE \cite{lin2004rouge}, METEOR \cite{meteor}) \cite{roy:fse2021,haque:icpc2022,mastropaolo:icse2024}, it seems reasonable to start considering the use of LLM as a judge in this context. Additional studies are needed to also investigate their complementarity with state-of-the-art metrics.
\section{Threats to Validity} \label{sec:threats}

{\bf Construct validity.} Using tests as a proxy for code correctness is a limitation of our study. To partially mitigate this issue, we performed preliminary checks aimed at excluding code generation problems clearly accompanied by an inadequate test suite. For the \CS study, we assessed the agreement among the human judges, showing the reliability of their judgement as an oracle.

{\bf Internal validity.} As for any work relying on manual analysis, there are possible subjectiveness issues in our findings. To alleviate such a threat, multiple evaluators were always involved in any manual step required in our paper. Another threat relates to the prompts used. We experimented with four different prompts for each task, with full results available in \cite{replication}. As discussed, while the prompts have an impact, they do not really affect the overall findings / lessons learned from our study.

{\bf External validity.} Although we experimented with several (8) LLMs, the generalizability of our findings is capped by (i) the two code-related tasks subject of the study and (ii) the focus on the Java and Python programming languages. Differentiated replications can help to corroborate/contradict our findings. 
\section{Related Work} \label{sec:related}

LLM-as-a-judge have become quite popular in the NLP community for the assessment of tools automating challenging generative problems. For example, Zheng \etal \cite{zheng_judging_2023} studied the usage of LLMs-as-a-judge to evaluate the capabilities of LLM-based chat assistants. The authors highlight four key limitations: (i) positional bias (\eg the LLM favors solutions in the first position when assessing a pool of different solutions); (ii) verbosity (\ie tendency to favor longer responses); (iii) self-enhancement bias (\ie the LLM favors answers generated by itself) and; (iv) limited reasoning ability (\ie the LLM experiences issues when assessing answers related to math/reasoning questions). Our study design has been influenced by these findings. Indeed, our decision to ask the LLMs to judge each solution in isolation rather than in a comparative setting (\eg providing the LLM with $n$ possible implementations of a given function asking it to judge all of them in one shot) aims at avoiding positional bias and limit verbosity bias, since the model only sees one candidate to judge. In addition, we investigated self-enhancement bias.

With the aim of addressing the limitations of LLM as a judge, Huang \etal \cite{huang_limitations_2024} explicitly fine-tuned LLM for judging tasks, showing that while some improvements can be observed, limitations still remain. These include loss in generalizability (\ie the model specializes for a specific judging task, thus being unable to perform other tasks), fairness, and scalability. Investigating with DL models specifically fine-tuned as judges for code-related tasks is part of our future research agenda.

There are four works mostly related to our study \cite{koutcheme_evaluating_2024,Weyssow:llm-as-judge-se,zhuo-2024-ice,tong-zhang-2024-codejudge}. Weyssow \etal \cite{Weyssow:llm-as-judge-se} propose the use of LLM-as-a-judge for evaluating a software-related task automated by other LLMs. The judgment aimed to evaluate the LLMs capable of implementing code by meeting specific non-functional requirements (\eg maximizing code readability). Unlike our work, their focus is not on evaluating the LLMs as a judge, but rather on exploiting them in their methodology. In fact, as the only data point that shows that LLMs as judges can work in this context, the authors show that different LLMs tend to agree on the fact that the solutions generated by the GPT-based models are superior to others, similar to what we observed.

Koutcheme \etal \cite{koutcheme_evaluating_2024} evaluate the effectiveness of LLMs in generating feedback for 57 programming assignments at the beginner level. Two analyses are performed. The first is a qualitative analysis in which annotators focused on revealing the main issues experienced by the models while judging (\eg hallucination by mentioning non-existent bugs). The second is a quantitative one in which other LLMs judge the feedback generated by each LLM. Unlike Koutcheme \etal, the quantitative analysis in our work does not rely on other LLMs, but on ``more reliable'' oracles such as tests (\CG) and human assessments (\CS). In addition, we focus on two different tasks and substantially larger datasets.

Zhuo \etal \cite{zhuo-2024-ice} present ICE-Score, an evaluation metric that exploits GPT-3.5-turbo as a judge for code implementations. The authors experiment with both judging code ``usefulness'' and code ``functional correctness''. The latter is basically the same \CG task as considered in our work. The authors ask ICE-Score to judge 20 implemented versions generated for each of the 164 coding problems featured in the HumanEval-X benchmark \cite{humaneval-x}. These 164 problems are available in 18 different languages, four of which are considered by the authors (Java, C++, Python, and JavaScript). Then, they analyze the correlation between the ICE-Score's judgements and the test outcome, reporting weak to moderate correlations, depending on the programming language and on whether the reference solution (\ie an example of correct implementation) was provided or not in the prompt asking the judgement. When not provided, the judging performance of the ICE-Score usually drops. Unlike Zhuo \etal \cite{zhuo-2024-ice}, we adopt a more challenging benchmark (\ie CoderEval \cite{yu2024codereval}) that features more complex coding problems compared to HumanEval-X. Indeed, the latter only includes simple and standalone functions (\ie functions invoking or accessing only built-in functions and standard libraries) which might be easier to judge, and only represent a small percentage of the real functions which can be found in open source projects \cite{yu2024codereval}. Also, we look at aspects such as self-bias which are ignored in \cite{zhuo-2024-ice}, we experiment with more recent LLMs (GPT-4) and with two code-related tasks (\CS on top of \CG).

Tong and Zhang present CodeJudge \cite{tong-zhang-2024-codejudge}, another application of GPT-3.5-as-a-judge for code correctness. The main novelty compared to the ICE-Score is a prompt that guides the LLM in performing ``slow thinking'' \cite{kahneman2011thinking}. The authors show that, thanks to such a prompt, CodeJudge outperforms ICE-Score. For this reason, in our study we adopt the prompt proposed in \cite{tong-zhang-2024-codejudge} showing, however, no major improvements in the LLMs' capabilities of assessing the correctness of code implementations. Note that our findings for \CG are aligned with those reported by Tong and Zhang \cite{tong-zhang-2024-codejudge}. In fact, CodeJudge has been tested on four different benchmarks, two of which (\ie APPS \cite{hendrycksapps2021} and BigCodeBench \cite{zhuo2024bigcodebench}) are similar in complexity to CoderEval. On these two (Python) benchmarks, the authors report an overall accuracy of 57.00\% (APPS) and 54.56\% (BigCodeBench), which is aligned with what we observed on Python (56\%). Unlike \cite{tong-zhang-2024-codejudge}, also in this case we look at additional aspects (such as self-bias) and an additional task (\CS) for which we had to manually build evaluation benchmarks.
\section{Conclusions and Future Work} \label{sec:conclusion}

We investigated the effectiveness of LLMs-as-a-judge for two code-related tasks, namely \CG and \CS. We considered LLMs having different sizes, ranging from $\sim$1B (DeepSeek Coder \cite{guo:2024}) to hundreds of billions (GPT-4-turbo \cite{chatgpt}). The judgement tasks focused on code correctness (\ie is a given function correct?) and on code summary quality. For code correctness, we used test results to assess the correctness of the judgment, while for summary quality we correlated the LLMs' and humans' judgments. Our findings show that ``small'' LLMs struggle in judging tasks, with GPT-4-turbo being the model that achieves the best results. Still, even GPT-4-turbo frequently fails in assessing code correctness, while being a reliable judge of code summary quality.

Our future work will focus on experimenting with small LLMs specifically fine-tuned for code-related judgment tasks and extending our study to additional code-related tasks (\eg bug-fixing).

\bibliographystyle{IEEEtranS}
\bibliography{biblio}

\end{document}